\documentclass[fleqn,usenatbib]{mnras}

\usepackage{amsmath}
\usepackage{amssymb}
\usepackage[T1]{fontenc}
\usepackage[dvipsnames]{xcolor}
\usepackage{graphicx}
\usepackage{mathptmx}
\usepackage{mathrsfs}
\usepackage{txfonts}

\newcommand{\ain}{a_{\rm in}}
\newcommand{\anu}{a_\nu}
\newcommand{\As}{A_{\rm s}}
\newcommand{\Ci}{{\mathrm{Ci}}}
\newcommand{\classcode}{{\sc class}}

\newcommand{\cosmicenu}{{\sc Cosmic-E$\nu$}}
\newcommand{\COut}{C_{\rm O}}
\newcommand{\dcb}{\delta_{\rm cb}}
\newcommand{\DDcb}{\Delta^2_{\rm cb}}

\newcommand{\DDnu}{\Delta^2_\nu}
\newcommand{\dm}{\delta_{\rm m}}
\newcommand{\dmDO}{\delta_{\rm m}^{\rm (DO)}}
\newcommand{\dmIO}{\delta_{\rm m}^{\rm (IO)}}
\newcommand{\dmNO}{\delta_{\rm m}^{\rm (NO)}}
\newcommand{\dnu}{\delta_\nu}
\newcommand{\enuii}{{\sc Cosmic-E$\nu$-II}}
\newcommand{\EQ}[1]{Eq.~(\ref{#1})}
\newcommand{\EQEQ}[2]{Eqs.~(\ref{#1},~\ref{#2})}
\newcommand{\EQS}[2]{Eqs.~(\ref{#1}-\ref{#2})}
\newcommand{\fastnuf}{{\sc fast}-$\nu f$}
\newcommand{\fehdm}{f_{\textrm{\sc ehdm}}}
\newcommand{\fftm}{{\sc FlowsForTheMasses}}
\newcommand{\fnu}{f_\nu}
\newcommand{\Hc}{{\mathcal H}}
\newcommand{\gsef}{\xi}  
\newcommand{\gvir}{g_{\rm v}}
\newcommand{\Hco}{{\mathcal H}_0}
\newcommand{\Iyc}{{\mathcal I}^{\rm (c)}}
\newcommand{\Iys}{{\mathcal I}^{\rm (s)}}
\newcommand{\kfs}{k_{\rm fs}}

\newcommand{\knr}{k_{\rm nr}}
\newcommand{\lvir}{\ell_{\rm v}}
\newcommand{\mehdm}{m_{\textrm{\sc ehdm}}}
\newcommand{\Mnu}{M_\nu}
\newcommand{\mnus}{m_{\nu,s}}
\newcommand{\muflr}{{\sc MuFLR}}
\newcommand{\MTiv}{{\sc Mira-Titan-IV}}

\newcommand{\nOut}{n_{\rm O}}
\newcommand{\Ns}{N_s}
\newcommand{\ns}{n_{\rm s}}

\newcommand{\Obo}{\Omega_{\mathrm{b},0}}
\newcommand{\Ocb}{\Omega_\mathrm{cb}}
\newcommand{\Ocbo}{\Omega_{\mathrm{cb},0}}

\newcommand{\Odo}{\Omega_{\mathrm{d},0}}
\newcommand{\Om}{\Omega_\mathrm{m}}

\newcommand{\Omo}{\Omega_{\mathrm{m},0}}
\newcommand{\On}{\Omega_\nu}

\newcommand{\Ono}{\Omega_{\nu,0}}

\newcommand{\Onao}{\Omega_{\nu,\alpha,0}}
\newcommand{\Oro}{\Omega_{{\rm r},0}}
\newcommand{\Qnu}{Q_\nu}
\newcommand{\Qnus}{Q_{\nu,s}}
\newcommand{\Rnu}{{\mathcal R}}
\newcommand{\Rvir}{R_{\rm v}}
\newcommand{\Si}{{\mathrm{Si}}}
\newcommand{\Sigc}{\Sigma^{\rm (c)}}
\newcommand{\Sigs}{\Sigma^{\rm (s)}}
\newcommand{\sinit}{s_{\rm in}}
\newcommand{\supp}{\sigma_\nu}
\newcommand{\tconf}{{\mathcal T}}
\newcommand{\Tehdmo}{T_{{\textrm{\sc ehdm}},0}}
\newcommand{\Tnuo}{T_{\nu,0}}
\newcommand{\Upec}{U^{\rm (pec)}}
\newcommand{\Uth}{U^{\rm (th)}}

\newcommand{\yin}{y_{\rm in}}

\title{Slow neutrinos: non-linearity and momentum-space emulation}
\author[Amol Upadhye, {\em et al.}]{%
  Amol Upadhye$^{1,2}$\thanks{E-mail: a.upadhye@ynu.edu.cn}
  and Yin Li$^3$
  \\
  $^1$~South-Western Institute for Astronomy Research, Yunnan University, Kunming, Yunnan 650500, People's Republic of China\\
  $^2$~Yunnan Key Laboratory of Survey Science, Yunnan University, Kunming, Yunnan 650500, People's Republic of China\\
  $^3$~Department of Strategic \& Advanced Interdisciplinary Research, Peng Cheng Laboratory, Shenzhen, Guangdong 518000,\\
  \phantom{$^3$~}People's Republic of China
}

\begin{document}
\label{firstpage}
\pagerange{\pageref{firstpage}--\pageref{lastpage}}

\maketitle

\begin{abstract}
  Recent cosmological bounds on the sum of neutrino masses, $\Mnu=\sum m_\nu$, are in tension with laboratory oscillation experiments, making cosmological tests of neutrino free-streaming imperative.  In order to study the scale-dependent clustering of massive neutrinos, we develop a fast linear response method, \fastnuf{}, applicable to neutrinos and other non-relativistic hot dark matter.  Using it as an accurate linear approximation to help us reduce the dynamic range of emulator training data, based upon a non-linear perturbation theory for massive neutrinos, we improve the emulator's accuracy at small $\Mnu$ and length scales by a factor of two.  We significantly sharpen its momentum resolution for the slowest neutrinos, which, despite their small mass fraction, dominate small-scale clustering.  Furthermore, we extend the emulator from the degenerate to the normal and inverted mass orderings.  Applying this new emulator, \enuii{}, to large halos in N-body simulations, we show that non-linear perturbation theory can reproduce the neutrino density profile in the halo outskirts, $2\Rvir \lesssim r \lesssim 10\Rvir$, to better than $10\%$.
\end{abstract}


\section{Introduction}
\label{sec:int}

Despite hopes of a definitive cosmological $\Mnu$ measurement, recent upper bounds $\Mnu \leq 53$~meV from the cosmic microwave background (CMB) and baryon acoustic oscillations (BAO) \citep{Planck:2018vyg,DESI:2024hhd,DESI:2025ejh,SPT-3G:2025zuh} are in tension with the long-established lower bound $\Mnu \geq 59$~meV from laboratory oscillation experiments \citep{DayaBay:2022orm,Esteban:2024eli}.  Even more troubling, the cosmological data seem to prefer a small-scale enhancement of matter clustering, the opposite of the free-streaming suppression predicted in massive neutrino models, though this may be driven by the preference of the CMB data for stronger gravitational lensing than predicted by General Relativity \citep{Green:2024xbb,Sailer:2025lxj,Jhaveri:2025neg}.

Tight $\Mnu$ bounds are largely due to measurements of the cosmic geometry, such as the DESI measurement of the redshift-dependent BAO standard ruler, which are relatively robust with respect to astrophysical systematics.  However, a different distance indicator, the supernova luminosity distance, prefers a higher matter density today, which is strongly correlated with a higher $\Mnu$ \citep{Brout:2022vxf,Rubin:2023jdq,DES:2024jxu,Efstathiou:2024xcq,Loverde:2024nfi}.  Distance indicators have significant parameter degeneracies, as they depend upon the Hubble parameter, the spatial curvature, and the equation of state of the dark energy.  Moreover, tensions between different distance indicators have persisted for over a decade, possibly indicating a non-standard cosmology \citep{Freedman:2024eph,Riess:2025chq,Freedman:2025nfr,Leauthaud:2025azz,Ishak:2025cay}, which would significantly weaken the soundness of $\Mnu$ bounds from distances.

The importance of $\Mnu$ as a fundamental particle physics parameter, and its ability to differentiate between the normal ordering (NO) and inverted ordering (IO) of neutrino masses, make further astrophysical constraints essential.  Neutrino free-streaming is a unique cosmological phenomenon distinct from distance indicators, allowing for independent cross-checks on $\Mnu$ measurements which will become increasingly powerful as the cosmological data improve.  Aside from its scale-dependent suppression of linear matter clustering, free-streaming gives rise to several potentially-observable effects in the non-linear regime, such as a contribution to scale-dependent halo bias \citep{LoVerde:2014pxa,Chiang:2017vuk,Chiang:2018laa}; differences in neutrino capture by halos in neutrino-rich vs. neutrino-poor regions \citep{Yu:2016yfe}; apparent neutrino ``wakes'' caused by the gravitational focusing of neutrinos by halos \citep{Zhu:2013tma,Zhu:2014qma,Inman:2015pfa,Ge:2023nnh,Nascimento:2023ezc}; and a unique parity-odd contribution to the halo angular momentum field \citep{Yu:2018llx}.  Searches for such effects and their backgrounds are progressing; see \cite{Ge:2024kac,Tang:2025lmm,Moon:2025bbe,Nascimento:2025bax}.

However, accurate theoretical computations of these effects require numerically expensive N-body simulations. Including neutrinos as separate particles in a simulation substantially increases its computational expense, as neutrinos' large thermal velocity dispersion forces us to consider their full six-dimensional phase space, as against the three-dimensional spatial distribution of cold dark matter (CDM) and baryons.  Further, neutrinos' small-scale clustering is severely contaminated by shot noise, particularly at low $\Mnu$.

For this reason, the simplest neutrino simulations include neutrinos purely as fluids linearly responding to the non-linear clustering of CDM and baryons.   Several neutrino linear response methods have been proposed, by \cite{Ringwald:2004np,Wong:2008ws,Ali-Haimoud:2012fzp,Dupuy:2013jaa,Dupuy:2014vea,Dupuy:2015ega,Chen:2020kxi,Chen:2020bdf,Ji:2022iji,Pierobon:2024kpw,Lee:2025zym}.  Multi-fluid linear response, as implemented in the \muflr{} code of \cite{Chen:2020bdf}, is the slowest, though it also tracks the neutrinos' momentum distribution, which impacts the detectability of the  neutrino background in terrestrial experiments such as KATRIN~\citep{KATRIN:2024cdt} and PTOLEMY~\citep{PTOLEMY:2019hkd}.

Our first new result is a significantly faster multi-fluid linear response, which we call \fastnuf{}.\footnote{We make a simple implementation of \fastnuf{} publicly available online at {\tt http://codeberg.org/upadhye/FASTnuf}~.}  Running in milliseconds on a desktop computer, it is based upon an exact solution for neutrino test particles in the Einstein-de Sitter cosmology, allowing for the efficient treatment of highly-oscillatory integrands arising from small-scale clustering.  Next, we apply \fastnuf{} to the emulation of the \fftm{} non-linear perturbation theory for massive neutrinos, building upon the \cosmicenu{} emulator of \cite{Upadhye:2023bgx}.

\cosmicenu{} faced a number of challenges.  Firstly, its equal-density-binning of the neutrino population meant that it inadequately sampled the low thermal momenta that dominate clustering at small masses and length scales, as shown by \cite{Upadhye:2024ypg}, leading to $50\%$ errors at $k=1~h/$Mpc.  While this appeared at first glance to be a simple matter of increasing the low-momentum sampling, that reference also demonstrated that low-momentum neutrinos lead to increasingly severe numerical instabilities at high $\Mnu$.  This left emulators with an unappealing choice between suboptimal accuracy at low masses, which are preferred by the data, and a discontinuity in the momentum sampling as a function of $\Mnu$.  With \fastnuf{}, we instead interpolate between the lowest-momentum neutrinos in our existing training set, as well as the Time-RG perturbation theory of \cite{Pietroni:2008jx,Lesgourgues:2009am}, upon which \fftm{} was based.  This method achieves a similar accuracy to the momentum sampling of \cite{Upadhye:2024ypg} without the high-$\Mnu$ numerical instabilities.

Secondly, \cosmicenu{} emulated ten different neutrino thermal momentum bins, providing a coarse sampling of the neutrino distribution function.  Since slower neutrinos, those with lower thermal velocities, are more easily captured by cosmic structure, and the detectability of the cosmic neutrino background in terrestrial experiments itself depends upon neutrino momenta,  the theoretical prediction of momentum-dependent neutrino clustering is useful.  Through \fastnuf{}, we substantially improve upon the momentum resolution of \cosmicenu{}, allowing us to explore in greater detail the phase space of the cosmic neutrino background.

Thirdly, \cosmicenu{} assumed a degenerate ordering (DO) of neutrino masses, that is, the equality of the masses of the three mass eigenstates.  While this is an accurate approximation at high $\Mnu$, recent data prefer low $\Mnu$, and may soon be able to distinguish between NO and IO masses.  A better theoretical understanding of non-linear clustering in these two hierarchies will be useful for cross-checking the preference of laboratory and cosmological experiments for NO neutrinos.  Using the effective hot dark matter formalism of \cite{Upadhye:2024ypg,Bayer:2020tko}, we extend the emulator to NO and IO neutrinos.  We call the improved \fastnuf{}-based emulator \enuii{},\footnote{We make the \enuii{} emulator source code publicly available online at {\tt http://codeberg.org/upadhye/Cosmic-Enu-II}~.} and we test its accuracy by comparing its power spectra to those of N-body simulations available in the literature.

Finally, we apply \enuii{} to the prediction of the neutrino density profile around large CDM+baryon (cb) halos, $M_{\rm cb} = 10^{15}~M_\odot/h$, in the Quijote simulations of \cite{Villaescusa-Navarro:2019bje}.  For $\Mnu=100$~meV, $200$~meV, and $400$~meV, we can accurately predict the density perturbation $\delta_\nu(r) = \rho_\nu(r)/{\bar \rho}_\nu - 1$ to better than $10\%$ in the halo outskirts, $2 \Rvir \leq r \leq 10 \Rvir$, with $\Rvir$ the virial radius.  Unfortunately, this coincides with the transition between the $1$-halo and $2$-halo regimes, where even our knowledge of the CDM clustering is limited, preventing us from making a concrete observable prediction.  However, \cite{Diemer:2014xya,More:2015ufa,More:2016vgs,Diemer:2017bwl,Diemer:2017uwt,ONeil:2020zjy,Fong:2020fuf,Fong:2022jml,Zhou:2023uhb,Gao:2023kdu} have advanced cosmology's theoretical understanding of the halo outskirts by leaps and bounds over the past $10$-$12$ years.  This gives  us some hope of future halo-based constraints upon $\Mnu$ and the mass ordering.

This article is organized as follows.  After providing some background in Sec.~\ref{sec:bkg}, we detail our \fastnuf{} linear response procedure in Sec.~\ref{sec:fnf}.  In Sec.~\ref{sec:Rnl} we study the enhancement of neutrinos' clustering by non-linear fluid dynamics.   Section~\ref{sec:emu} applies this to the development of the \enuii{} emulator.  The neutrino density profile outside halos is computed in Sec.~\ref{sec:halo}, and Sec.~\ref{sec:con} concludes.

\section{Background}
\label{sec:bkg}

\subsection{Exact solutions for {\em w$_0$}CDM models}
\label{sub:bkg:exact}

The \cosmicenu{} emulator, and the \MTiv{} emulator upon whose cb clustering it was based, use the \cite{Chevallier:2000qy,Linder:2002et} (CPL) parameterization of the dark energy equation of state,
\begin{equation}
  w(z) = w_0 + w_a z / (1+z),
  \label{e:w_CPL}
\end{equation}
which smoothly transitions from $w_0$ at $z=0$ to $w_0+w_a$ as $z \rightarrow \infty$.  The resulting conformal Hubble rate $\Hc = a H$ for spatially-flat models after radiation has become negligible is:
\begin{equation}
  \Hc(z) =
  \Hco \sqrt{\Omo \,\,(1+z) + \Odo \,\,(1+z)^{1+3(w_0+w_a)} e^{-3w_a z/(1+z)} }
  \label{e:w0waCDM_Hc}
\end{equation}
where $\Hco = 100 h~/~{\rm Mpc}~/~299792.458$ in units where the speed of light is unity, and $\Omo$ and $\Odo$ are, respectively, the density fractions of matter and dark energy today.

We know of no exact conformal time and growth factor solutions for the general CPL model, but such solutions do exist for constant-$w$ dark energy parameterizations, with $w_a=0$, which we refer to as $w_0$CDM models.
Defining $r_0 := \Omo/\Odo$, we find the following expressions for the physical time $t(a)$, the conformal time $\tconf(a)$, and the superconformal time $s(a)$, with the latter two defined by $d\tconf = dt/a$ and $ds = dt/a^2$:
\begin{eqnarray}
  \tfrac{3\Hco \Omo^{1/2} t}{2a^{3/2}}
  &=&
  (1 \!+\! \tfrac{a^{-3w_0}}{r_0})^{-\frac{1}{2}}
  {}_2F_1\left[\tfrac{1}{2}, 1, 1 \!-\! \tfrac{1}{2w_0},
    \tfrac{a^{-3w_0}/r_0}{1+a^{-3w_0}/r_0}\right],
  \label{e:w0CDM_t}
  \\
  \tfrac{\Hco \Omo^{1/2} \tconf}{2a^{1/2}}
  &=&
  (1 \!+\! \tfrac{a^{-3w_0}}{r_0})^{-\frac{1}{6w_0}}
  {}_2F_1\left[-\tfrac{1}{6w_0},
    \tfrac{1}{2} \!-\! \tfrac{1}{6w_0},
    1 \!-\! \tfrac{1}{6w_0},
    \tfrac{a^{-3w_0}/r_0}{1+a^{-3w_0}/r_0}\right],
  \label{e:w0CDM_tconf}
  \\
  \tfrac{\Hco \Omo^{1/2} s}{2a^{-1/2}}
  &=&
  -(1 \!+\! \tfrac{a^{-3w_0}}{r_0})^{-\frac{1}{6w_0}}
  {}_2F_1\left[\tfrac{1}{6w_0},
    \tfrac{1}{2} \!+\! \tfrac{1}{6w_0},
    1 \!+\! \tfrac{1}{6w_0},
    \tfrac{a^{-3w_0}/r_0}{1+a^{-3w_0}/r_0}\right].~\quad
  \label{e:w0CDM_s}
\end{eqnarray}
The comoving distance $\chi(a)$ to scale factor $a$ is the conformal lookback time $\tconf(1)-\tconf(a)$, with $a=1$ today, hence it may be found from \EQ{e:w0CDM_tconf}.  At small $a$, the $w_0$CDM cosmologies approach the Einstein-de Sitter (EdS) cosmology, therefore:  $t \propto a^{3/2}$; $\tconf \propto a^{1/2}$; and $s$ diverges, $s \propto -a^{-1/2}$.  For $w_0=-1$, \EQ{e:w0CDM_t} simplifies to $\tfrac{3}{2}\Hco \Odo^{1/2} t = \ln(\sqrt{a^3/r_0}+\sqrt{1+a^3/r_0})$.

The general solution to the second-order differential equation for the linear matter density contrast $\delta_{\rm m}$ has been published in \cite{Lee:2009gb,Lee:2009if}. Choosing the growing mode $D(a)$ and normalizing it to approach $a$ at early times, we find:%
\begin{eqnarray}
  \tfrac{D}{a}
  &=&
  (1 \!+\! \tfrac{a^{-3w_0}}{r_0})^{\frac{1}{3w_0}}
  {}_2F_1\left[-\tfrac{1}{3w_0},
    \tfrac{1}{2} \!-\! \tfrac{1}{3w_0},
    1 \!-\! \tfrac{5}{6w_0},
    \tfrac{a^{-3w_0}/r_0}{1+a^{-3w_0}/r_0}\right]
  \label{e:w0CDM_D}
  \\
  \tfrac{dD}{da}
  &=&
  (1 \!+\! \tfrac{a^{-3w_0}}{r_0})^{\frac{1}{3w_0}\!-\!1}
  \!{}_2F_1\left[-\tfrac{1}{3w_0},
    \tfrac{1}{2} \!-\! \tfrac{1}{3w_0},
    1 \!-\! \tfrac{5}{6w_0},
    \tfrac{a^{-3w_0}/r_0}{1+a^{-3w_0}/r_0}\right]
  \nonumber\\
  &~&
  +\tfrac{2-3w_0}{5-6w_0} \tfrac{a^{-3w_0}}{r_0}
  (1 \!+\! \tfrac{a^{-3w_0}}{r_0})^{\frac{1}{3w_0}\!-\!2}
  \nonumber\\
  &~&\quad
  \times ~{}_2F_1\left[1\!-\!\tfrac{1}{3w_0},
    \tfrac{3}{2} \!-\! \tfrac{1}{3w_0},
    2 \!-\! \tfrac{5}{6w_0},
    \tfrac{a^{-3w_0}/r_0}{1+a^{-3w_0}/r_0}\right].
  \label{e:w0CDM_dDda}
\end{eqnarray}
Incidentally, the above formulae illustrate the difficulty in applying EdS initial conditions ($D=a$, $dD/da=1$) to numerical integration of the growth factor for dark energy models with small, negative $w$ at early times.  At sufficiently low $a$, the hypergeometric functions are approximately unity, so any deviation from the EdS growth is due to the factor $(1+a^{-3w}/r_0)^{1/(3w)}$.  At small $a$, requiring that the fractional deviation of this quantity from unity be less than $\epsilon$ implies $a<(3 |w| r_0 \epsilon)^{1/\left|3w\right|}$, which is $\sim 10^{-128}$ for $w = -0.01$ and $\epsilon=0.01$.

In the interests of numerical efficiency, we apply \EQS{e:w0CDM_tconf}{e:w0CDM_dDda} whenever $w_a$ is negligible.  In the general CPL case, we apply the growth factor integration procedure of \cite{Linder:2003dr}, starting from an initial scale factor of $10^{-150}$.

\subsection{Linear growth approximations}
\label{sub:bkg:growth}

Our work is built upon the multi-fluid perturbation theory for neutrinos whose foundations were laid by \cite{Dupuy:2013jaa,Dupuy:2014vea,Dupuy:2015ega}, and which was integrated into N-body simulations and developed into a non-linear perturbation theory by \cite{Chen:2020bdf,Chen:2022dsv,Chen:2022cgw,Upadhye:2023bgx,Pierobon:2024kpw,Upadhye:2024ypg}.  An alternative to the standard approach of \cite{Bond:1983hb,Ma:1995ey}, multi-fluid perturbation theory splits the four-velocity of each neutrino particle into two components: a thermal part, $\Uth_\mu(a)$, which is spatially homogeneous and whose lower-index spatial components are constant in time after neutrino decoupling; and a peculiar part, $\Upec_\mu(a,\vec x)$, which is sourced by the gravitational potentials.

Defining this constant spatial part to be%
\begin{eqnarray}
  \vec u &:=& \sqrt{(\Uth_1)^2 + (\Uth_2)^2 + (\Uth_3)^2} \\
 \Rightarrow \Uth_0(a) &=& -\sqrt{u^2 + a^2}
\end{eqnarray}
with $u := |\vec u|$, we treat neutrinos of mass $m_\nu$ after decoupling as described by a Fermi-Dirac distribution function, $f(m_\nu \vec u) = [\exp(m_\nu u / \Tnuo) + 1]^{-1}$, with $\Tnuo \approx (4/11)^{1/3} (3.044/3)^{1/4}T_{\gamma,0} = 1.9525$~K, and we approximate the $N_{\rm eff}=3.044$ effective neutrino species found by \cite{Bennett:2019ewm,Bennett:2020zkv,EscuderoAbenza:2020cmq,Akita:2020szl,Froustey:2020mcq} as fully thermalized.  Meanwhile, the peculiar velocity $\Upec_\mu$ vanishes in the subhorizon limit at early times.  Thus each neutrino sub-population described by a given $\vec u$ can be treated as a separate fluid.  This represents a Lagrangian description of the neutrinos in momentum space, since neutrinos are specified entirely by their initial four-velocity $(\Uth_0(\ain),\vec u)^T$, with no crossing from $\vec u$ to $\vec u' \neq \vec u$.

Multi-fluid perturbation theory treats each sub-population defined by $\vec u$ using its own set of fluid equations, with all sub-populations coupled only through their contributions to the gravitational potentials.  Since the direction of $\vec u$ enters into the fluid equations only through its angle with the Fourier vector, whose cosine $\mu = \vec u \cdot \vec k /(uk)$, different $\vec u$ with the same magnitude $u$ obey the same fluid equations.  We refer to all such fluids as a ``flow'' specified by the magnitude $u$.

We work in the sub-horizon, non-relativistic approximation, where the thermal component of the $3$-velocity is $u/a$.  The fluid equations for the dimensionless density contrast $\delta(a,\vec x,u,\mu) = \rho(a,\vec x, u,\mu) / {\bar \rho}(a,u) - 1$ and inflow $\theta(a,\vec x,u,\mu) := -\vec\nabla \cdot \vec V / \Hc$ of the peculiar velocity $V^i(a,\vec x,u,\mu) = U^{{\rm (pec)}\,i} / a$, are, in Fourier space,
\begin{eqnarray}
  \frac{\partial \delta}{\partial \ln a}
  &=&
  -\frac{ik\mu u}{a \Hc} \delta + \theta
  \label{e:eom_delta}
  \\
  \frac{\partial \theta}{\partial \ln a}
  &=&
  -\left(1+\frac{d \ln \Hc}{d \ln a}\right)
  -\frac{ik\mu u}{a \Hc} \theta - \frac{k^2}{\Hc^2}\Phi.
  \label{e:eom_theta}
  \\
  k^2 \Phi
  &=&
  -\frac{3}{2} \Hc^2 \Om \dm
  = -\frac{3}{2} \Hc^2 [\Ocb \dcb + \On \dnu]
  \label{e:eom_Poisson}
  \\
  \dnu
  &=&
  \frac{\int u^2 du  f(m_\nu u) \int \!\tfrac{d\mu}{2} \delta(a,k,u,\mu)
  }{\int u^2 du f(m_\nu u) }
  \!=\!
  \frac{\int u^2 du  f(m_\nu u) \delta(a,k,u)}{\int u^2 du f(m_\nu u) }\quad~~
  \label{e:def_dnu}
\end{eqnarray}
Here, the gravitational potential $\Phi$ depends through Poisson's \EQ{e:eom_Poisson} upon the total matter density $\dm(a,k)$, which, in turn, depends upon the neutrino flow monopoles $\delta(a,k,u) = \int_{-1}^1 \tfrac{d \mu}{2} \delta(a,k,u,\mu)$.  Henceforth, we use $\delta(a,k,u)$ to denote the monopole component of the solution to \EQS{e:eom_delta}{e:def_dnu}, and $\delta(a,k,u,\mu)$ for the full $\mu$-dependent solution. The \muflr{} (Multi-Fluid Linear Response) code of \cite{Chen:2020bdf} directly solves this system of equations, coupled to a perturbative linear or Time-RG $\dcb$, and that reference thoroughly discusses these solutions.

Neutrino clustering splits neatly into two regimes, the long-wavelength ``clustering'' regime and the short-wavelength ``free-streaming'' regime, above and below a characteristic free-streaming length.  For a given flow $u$ in Fourier space, the corresponding free-streaming wave number is%
\begin{equation}
  \kfs(a,u)
  = \sqrt{ \frac{3 \Om(a) \Hc(a)^2 a^2}{2 u^2 } }
  = \sqrt{ \frac{3 \Omo \Hco^2 a}{2 u^2} }.
  \label{e:kfs_a_u}
\end{equation}
The density contrast $\delta(a,k,u)$ of a given flow $u$ clusters just like the total matter density $\dm$ for $k \ll \kfs(a,u)$, and $\delta(a,k,u) / \dm \rightarrow \kfs(a,u)^2/k^2$ for $k \gg \kfs(a,u)$, leading \cite{Ringwald:2004np,Wong:2008ws,Chen:2020kxi,Pierobon:2024kpw} to suggest the approximations
\begin{eqnarray}
  \frac{\delta(a,k,u)}{\dm(a,k)}
  &\approx&
  \frac{1}{[1 + k/\kfs(a,u)]^2}
  =:
  \gsef(a,k,u)\quad
  \label{e:gsef_a_k_u}
  \\
  \frac{\dnu(a,k)}{\dm(a,k)}
  &\approx&
  \gsef(a,k,c_\nu)
  ~\mathrm{where}~
  c_\nu^2 := \frac{3 \zeta(3) \Tnuo^2 }{ 2 \log(2) m_\nu^2 }
  \label{e:gsef_a_k_cnu}
\end{eqnarray}
for flow $u$ and the neutrino population as a whole, respectively.  Note that \EQ{e:gsef_a_k_cnu} does not follow from \EQ{e:gsef_a_k_u}, but both interpolate between the same clustering and free-streaming limits, giving similar results for $\dnu$.  Here, $c_\nu$ is a sound speed for the neutrino population.

Next, we consider the impact of neutrinos on linear cb growth.  \cite{Bond:1980ha,Hu:1997vi} show that, in the EdS cosmology, neutrino masses corresponding to a fraction $\fnu = \Ono / \Omo$ of the total matter reduce the high-$k$ growth factor from $a$ to $a^{1-\Qnu}$ well after neutrinos go non-relativistic at scale factor $\anu$:%
\begin{eqnarray}
  \lim_{k \rightarrow \infty}\frac{D_{\rm cb}(a,k,m_\nu)}{D_{\rm cb}(a,k,0)}
  &=& \frac{(a/\anu)^{1-\Qnu}}{(a/\anu)}
  \approx 1-\Qnu \log\left(\frac{a}{\anu}\right)
  \label{e:nu_supp_EdS}
  \\
  \mathrm{where}~
  \Qnu &=& \frac{5}{4} \left(1 - \sqrt{1 - \frac{24}{25}\fnu}\right),
  \quad
  \anu = \frac{7\pi^4 \Tnuo}{180\zeta(3)m_\nu}.\qquad
  \label{e:def_Qnu_anu}
\end{eqnarray}
Approximately, $\anu = p_{\nu,0}/m_\nu \approx 3.15 \Tnuo/m_\nu$, and for $\fnu \ll 1$ we have $\Qnu \approx \frac{3}{5} \fnu$.  Since EdS is a reasonable approximation to the universe at the time when neutrinos go non-relativistic, and remains so until the recent dark energy domination, this estimate closely approximates small-scale neutrino suppression.  For $a = 1$, and typical neutrino density fractions, \EQ{e:nu_supp_EdS} reduces to $\approx 1-3\fnu$, implying a suppression of $\approx 6 \fnu$ in the cb power and $8 \fnu$ in the matter power.

Our remaining task is to approximate its scale-dependence.  The largest scale affected by this suppression is approximated by setting $a=\anu$ and $u=c_\nu$ in \EQ{e:kfs_a_u}:%
\begin{equation}
  \knr := \kfs(\anu,c_\nu)
  = \sqrt{\frac{7\pi^4 \log(2) \Omo \Hco^2 m_\nu}{180\zeta(3)^2 \Tnuo}}.
  \label{e:def_knr}
\end{equation}
This is approximately $1.35 \Hco \sqrt{\Omo m_\nu / \Tnuo}$, and is $\ll \kfs(1,c_\nu)$, which scales as an extra power of the large quantity $\anu^{-1/2} \propto \sqrt{m_\nu/\Tnuo}$.
Following \cite{Ringwald:2004np}, we seek a low-order rational function in $k/\knr$ to generalize \EQ{e:nu_supp_EdS} to arbitrary $k$.  Using a least-squares fit for log-spaced $k$ values at $z=0$, for the $101$ massive-neutrino models of \cite{Moran:2022iwe}, we find
\begin{eqnarray}
  \supp(a,k)
  &:=&
  \frac{D_{\rm cb}(a,k,m_\nu)}{D_{\rm cb}(a,k,0)}
  \approx
  1 -\frac{\Qnu \log(a/\anu) k^2}{\beta_0 \knr^2 + \beta_1 k \knr + k^2}
  \label{e:supp_nu}
  \\
  \mathrm{with}~
  \beta_0 &=& 5.5
  ~\mathrm{and}~
  \beta_1 = 1.8
\end{eqnarray}
to fit with an RMS error of $0.6\%$ at $z=0$, and $0.9\%$ for $z\leq3$.

The growth factor of \EQ{e:w0CDM_D}, and its generalization to time-dependent dark energy equations of state such as \EQ{e:w_CPL}, apply to models containing only cold matter and dark energy.  At earlier times, the presence of radiation cannot be neglected.  The subhorizon growing mode solution for a universe containing only matter and radiation is $D(a) = a + 2 a_{\rm mr}/3$, where $a_{\rm mr} = \Oro/\Omo$ and $\Oro$ is the radiation density fraction today.  Thus our final approximation for the linear growth factor of a flow $u$ is
\begin{equation}
  D(a,k,u)
  \approx
  \left[ D(a) + \frac{2}{3}a_{\rm mr} \right]
  \supp(a,k) \gsef(a,k,u)
  \label{e:D_a_k_u}
\end{equation}
with $D(a)$ given by \EQ{e:w0CDM_D} or its generalization to a varying equation of state; $\supp$ given by \EQ{e:supp_nu}; and $\gsef$ given by \EQ{e:gsef_a_k_u}. Since $\kfs(a,u)\rightarrow\infty$ and $\gsef(a,k,u)\rightarrow 1$ as $u\rightarrow 0^+$, \EQ{e:D_a_k_u} in the zero-velocity limit reduces to the cb growth factor corrected for the effects of early-time radiation and late-time, small-scale neutrino growth suppression.  This is expected, as the cold limit of HDM is CDM.

\subsection{Neutrino mass ordering}
\label{sub:bkg:mass_order}

Neutrino oscillations are described by six independent parameters: a CP-violating phase angle, three mixing angles, and two mass splittings, defined as $\Delta m_{ij}^2 := m_{\nu,i}^2 - m_{\nu,j}^2$.  The NuFit-6.0 joint analysis of oscillation data, \cite{Esteban:2024eli}, provides up-to-date constraints on each of these, though only the mass splittings are relevant here.  The smaller-magnitude mass splitting $\Delta m_{21}^2 = 74.9$~meV$^2$ has a $3\sigma$ precision of $15\%$.

The larger-magnitude mass splitting may take either sign.  In the normal mass ordering it is positive, implying $m_{\nu,1} < m_{\nu,2} < m_{\nu,3}$, and \cite{Esteban:2024eli} finds $\Delta m_{31}^2 = 2513$~meV$^2$ with a relative precision of $5.1\%$.  In the inverted ordering it is negative, with $m_{\nu,3} < m_{\nu,1} < m_{\nu,2}$, and \cite{Esteban:2024eli} finds $\Delta m_{31}^2 = -2484$~meV$^2$, with the same relative precision.

Since these experimental undertainties are much smaller than cosmological uncertainties in the neutrino masses, and will continue to be for the forseeable future, we neglect them and fix the mass splittings at their best-fit values.  They imply minimum $\Mnu$ of $58.8$~meV (NO) and $98.9$~meV (IO).

\subsection{Non-linear effective hot dark matter}
\label{sub:bkg:ehdm}

The multi-fluid perturbation theory discussed in Sec.~\ref{sub:bkg:growth} begins by discretizing the neutrinos' Fermi-Dirac distribution function.  Of course, this procedure may be applied to any other distribution function $f$.  \cite{Bayer:2020tko,Upadhye:2024ypg} demonstrated that an arbitrary collection of hot dark matter (HDM) species could be combined into a single effective hot dark matter (EHDM) species with an appropriately-defined effective distribution function.

The starting point is the collisionless Boltzmann equation for a massive particle, written in terms of the four-velocity $U_\alpha$ rather than the four-momentum $P_\alpha = m U_\alpha$:%
\begin{equation}
  U^0 \frac{\partial f}{\partial \tconf}
  + U^i \frac{\partial f}{\partial x^i}
  - \Gamma^i_{\alpha\beta} U^\alpha U^\beta \frac{\partial f}{\partial U^i}
  = 0.
  \label{e:collisionless_Boltzmann_vs_U}
\end{equation}
The $\Gamma^\gamma_{\alpha\beta}$ are Christoffel symbols.  Since the individual HDM mass does not appear, a manifestation of the equivalence principle, the distributions of all such species will evolve by exactly the same fraction, provided that all non-gravitational interactions have decoupled.  Thus we may define a single EHDM species, of any mass $\mehdm$ and temperature constant $\Tehdmo$, whose distribution function is
\begin{equation}
  \fehdm(\mehdm U_\alpha,x_\beta)
  :=
  \!\sum_S \!\frac{g_S m_S^4}{\mehdm^4} \!f_S(m_S U_\alpha,x_\beta)
  \label{e:def_f_ehdm}
\end{equation}
for which the stress-energy tensor $T_{\mu\nu}$ exactly matches the sum of individual HDM stress-energy tensors.  Here, the $S$ refers to the individual HDM species, with multiplicities $g_S$, masses $m_S$, temperature constants $T_{S,0}$, and distribution functions $f_S(P_\alpha,x_\beta)$. The mass-independence of \EQ{e:collisionless_Boltzmann_vs_U} ensures that this continues to represent all of the HDM their non-gravitational interactions are negligible.

Combined with multi-fluid perturbation theory, this represents a significant improvement in efficiency, particularly when the $T_{S,0} / m_S$ for the different species are not significantly different.  As \cite{Upadhye:2024ypg} shows, $10$-$20$ flows are adequate for computing the non-linear clustering of the total EHDM representing three standard neutrino species, as well as recovering the individual-species power spectra.  Here, we focus on neutrinos alone, and we follow that reference in defining $\mehdm=\Mnu/3$ and $\Tehdmo = \Tnuo$.

\section{{\pmb \fastnuf{}} procedure}
\label{sec:fnf}

\subsection{Exact Einstein-de Sitter test-$\nu$ solution}
\label{sub:fnf:EdS}

The linear density monopole of a non-relativistic neutrino flow specified by $u$ is given in \cite{Chen:2022dsv} by
\begin{eqnarray}
  \delta(a,k,u)
  &=&
  \frac{3 \Om(a) \Hc(a)^2 a}{2 u k}
  \!\!\!\int_{\sinit}^s \!ds' \, \sin\left(u k (s-s')\right) a(s') \dm(a(s'),k)
  \nonumber
  \\
  &=&
  \frac{3\Omo \Hco^2}{2 u^2 k^2}
  \int_{y(a)}^{\yin} dy' \sin(y'-y) a(y') \dm(a(y'),k),
  \label{e:d0_a_k_u_formal}
  \\
  \mathrm{with}~
  y(a)
  &:=&
  -u k s(a),
  \quad
  \yin = y(\ain),
  \quad
  \sinit = s(\ain).
\end{eqnarray}
Note that $y$ has been defined so as to be positive for all $0 \leq a \leq 1$, for which \EQ{e:w0CDM_s} implies $s<0$; thus, the distant past corresponds to large, positive $y$.  Equation~(\ref{e:d0_a_k_u_formal}) is merely a formal solution, since the total neutrino density made up of the $\delta(a,k,u)$ enters into the total matter density $\dm(a,k)$ on the right.  However, since neutrinos are only $\sim 1\%$ of the total matter, $\dm(s,k)$ can be obtained accurately from just a few flows, and then \EQ{e:d0_a_k_u_formal} used to improve the phase-space resolution of the neutrino density.

But first, we integrate \EQ{e:d0_a_k_u_formal} exactly in an Einstein-de Sitter universe in which neutrinos may be treated purely as test particles making up a negligible fraction of the universe's energy density, hence contributing nothing to the gravitational potential.  We refer to this as the EdS-test-$\nu$ case.  In this case, $\dm(a,k) = a \dm(1,k)$, and $a^{1/2} = -2/(\Hco s) \propto 1/y$, as may be seen by taking the $r_0\rightarrow\infty$ limit of \EQ{e:w0CDM_s}.

The result is that the integrand of \EQ{e:d0_a_k_u_formal} is proportional to $\sin(y'-y) / (y')^4$, so the integral may be evaluated exactly.  The most useful case is the limit $\ain \rightarrow 0^+$, corresponding to $\yin \rightarrow \infty$, for which we find
\begin{equation}
  \left. \frac{\delta(a,k,u)}{\dm(a,k)}\right|_{\mathrm{EdS,test-}\nu}
  =
  1 + y^2 \cos(y) \Ci(y) + y^2\sin(y)\left[\Si(y)-\frac{\pi}{2}\right].
\end{equation}
A further simplification arises from noting that, for the EdS case alone, $y(a) = \sqrt{6} k / \kfs(a,u)$.  The clustering limit $\delta(a,k,u)\rightarrow\dm(a,k)$ as $k\rightarrow 0$ is readily apparent, while the free-streaming limit $\delta(a,k,u)\rightarrow \kfs(a,k)^2 \dm(a,k) / k^2$ as $k\rightarrow \infty$ can be seen from the asymptotic behaviors of $\Ci(y)$ and $\Si(y)$.

\subsection{General cosmologies}
\label{sub:fnf:gen_cos}

Equation~(\ref{e:d0_a_k_u_formal}) can be integrated exactly in the EdS-test-$\nu$ case because $a(y') \dm(a(y'),k) \propto a(y')^2 \propto (y')^{-4}$, and the resulting integrand $\propto \sin(y'-y)/(y')^4$ is exactly integrable over $y'$ for any real $y$.  The main result of this Section begins with the observation that $\sin(y'-y)/(y')^n$ may be integrated exactly for any integer $n$, as may be seen for $n>0$ by expanding $\sin(y'-y)=\sin(y')\cos(y)-\cos(y')\sin(y)$ and then integrating repeatedly by parts.  Thus, we do not need a closed-form expression for the time-dependence of $a \dm$ at a given wave number; we need only interpolate it using polynomials in $s$.  Dividing the interval $[\sinit,s]$ into tens of sub-intervals, we may easily approximate $a \dm$ on each one using a cubic spline or other low-order interpolating polynomial.  The advantage is apparent from an examination of the integrand of \EQ{e:d0_a_k_u_formal} at high $k$, whose highly-oscillatory nature  would appear to require many more than $k|s| \sim 2k/\Hco \sim 10^4 k[h/{\rm Mpc}]$ time steps.

Begin by defining the $s$-dependent function
\begin{equation}
  g(s,k) := (\Hco s)^4 a(s) \, \dm(a(s),k)
  \label{e:def_g_s_k}
\end{equation}
for any cosmology.  We may choose any integer power $N_g$ of $\Hco s$ in \EQ{e:def_g_s_k}; we pick $N_g=4$ because, in the EdS-test-$\nu$ case, $g$ is constant.  The integrand of \EQ{e:d0_a_k_u_formal} is proportional to $\sin(uk(s-s')) g(s',k) / (s')^4$.  Let $[\sinit, s(a\!\!=\!\!1)]$ be divided into $\Ns-1$ sub-intervals $[s_j,s_{j+1}]$, with $s_0:=\sinit < s_1 < \ldots < s_{\Ns-1}=s(a\!\!=\!\!1)$, such that $g(s,k)$ at any $k$ may be approximated as a low-order polynomial on each sub-interval.  Substituting this polynomial into the integrand turns it into a series of terms $\propto \sin(y'-y)/(y')^n$ for different integers $n$, with no other dependence upon $y'$, allowing \EQ{e:d0_a_k_u_formal} to be integrated on each interval with no further approximation.

We limit ourselves here to cubic polynomials, $N_g-1=3$, though our results generalize to any order.  On the $j$-th superconformal time sub-interval $[s_j,s_{j+1}]$, let $g(s,k)$ be approximated $\sum_{n=0}^{N_g-1} g_{nj}(k) (\Hco s)^n$, and expand $\sin(uk(s-s')) = \sin(uks)\cos(uks')-\cos(uks)\sin(uks')$.  Defining
\begin{equation}
  \Iys_n(y) := \int_y^\infty dy' \frac{\cos(y')}{(y')^n}
  ~~\mathrm{and}~~
  \Iyc_n(y) := \int_0^y dy' \frac{\sin(y')}{(y')^n}
  \label{e:def_Iys_Iyc}
\end{equation}
as well as letting $\Iys_{nj}=\Iys_n(y_j)$ and $\Iyc_{nj}=\Iyc_n(y_j)$, we may compute all necessary integrals, as in Appendix~\ref{app:gcubic}.  The integral over the $j$-th sub-interval is $\sum_n g_{nj}(k) (\tfrac{\Hco}{uk})^n [\sin(y) (\Iys_{4-n,j}-\Iys_{4-n,j+1}) + \cos(y) (\Iyc_{4-n,j}-\Iyc_{4-n,j+1})]$.  All dependence upon the external variable $a$, through $s(a)$ and $y(a)$, has been pulled outside of the integral.

Collecting results, our solution to \EQ{e:d0_a_k_u_formal} for arbitrary $a$-dependent $\dm$, at scale factor $a_j = a(s_j)$ and wave number $k$, for flow $u$, with $y_j = -u k s_j$, is%
\begin{eqnarray}
  \delta(a_j,k,u)
  &=&
  \frac{3}{2}\Omo
  \left[\sin(y_j)\Sigs_j(k,u) + \cos(y_j)\Sigc_j(k,u)\right],
  \label{e:d0_a_k_u_fastnf}
  \\
  \Sigs_{0}
  &=&
  -g(\sinit,k) \left(\frac{u k}{\Hco}\right)^{\!2}
  \left(\Iys_{4,0} - \frac{\pi}{12}\right),
  \label{e:Sigs_j0}
  \\
  \Sigc_{0}
  &=&
  -g(\sinit,k) \left(\frac{u k}{\Hco}\right)^{\!2} \Iyc_{4,0},
  \label{e:Sigc_j0}
  \\
  \Sigs_{j+1} &=& \Sigs_{j}
  + \!\! \sum_{n=0}^{N_g-1} (-1)^n g_{nj}(k) \left(\frac{u k}{\Hco}\right)^{\!2-n}
  \left[ \Iys_{4-n,j} - \Iys_{4-n,j+1} \right],\qquad
  \label{e:Sigs}
  \\
  \Sigc_{j+1} &=& \Sigc_{j}
  + \!\! \sum_{n=0}^{N_g-1} (-1)^n g_{nj}(k) \left(\frac{u k}{\Hco}\right)^{\!2-n}
  \left[ \Iyc_{4-n,j} - \Iyc_{4-n,j+1} \right].
  \label{e:Sigc}
\end{eqnarray}
We point out three details.  Firstly, \EQS{e:Sigs_j0}{e:Sigc_j0} reflect our choice to extrapolate back to $a=0$, hence $s\rightarrow -\infty$, by assuming an EdS universe for $s \leq \sinit$.  It is precisely in this cosmology that $g(s,k)$ becomes independent of $s$ and can be evaluated at $\sinit$.  Secondly, by defining $\Sigs_j$ and $\Sigc_j$ as in \EQS{e:Sigs}{e:Sigc}, we may use \EQ{e:d0_a_k_u_fastnf} at any sub-interval endpoint $a_j = a(s_j)$, without further computation.  That is, although the integrand of \EQ{e:d0_a_k_u_formal} depends upon the time $s(a)$ at which $\delta(a,k,u)$ is evaluated, this dependence is factored outside of the integral in \EQ{e:d0_a_k_u_fastnf}.  Given a predetermined set of output times, we need only include these among the sub-interval endpoints.  Thirdly, the chief computational expense of this method, if fine samplings of $k$ and $u$ are required, is the need to compute $\Si(y)$, $\Ci(y)$, $\sin(y)$, and $\cos(y)$ over a large number of $u$, $k$, and $|s|$.  This may be mitigated, for example, through logarithmic sampling of each of these quantities with the same logarithmic step size, allowing the same $y$ to be reused for multiple combinations of $u$, $k$, and $s$.

\subsection{Tests of linear response}
\label{sub:fnf:tests}

\begin{figure}
  \hskip-3mm\includegraphics[width=92mm]{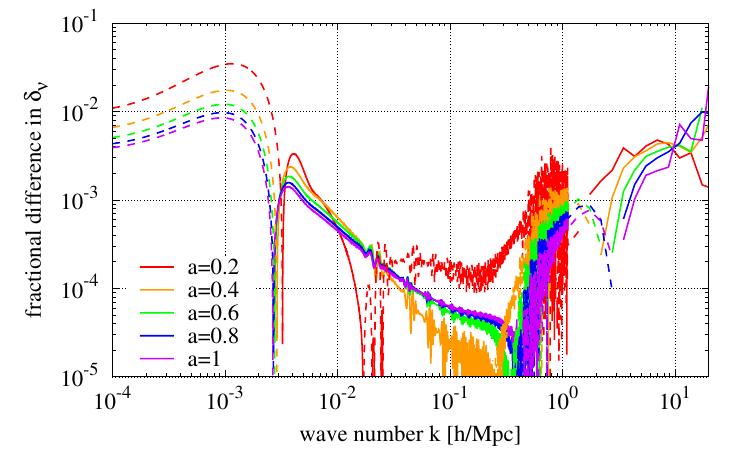}%
  \caption{
    Accuracy of \fastnuf{} for the total neutrino density contrast $\dnu(a,k)$.
    Shown is the fractional difference
    between \fastnuf{} and \classcode{} computations.  Dashed lines denote negative values,
    $\dnu^{({\rm FAST}\nu f)} < \dnu^{\rm (CLASS)}$.
    The two agree to better than $1\%$ for all $a \geq 0.2$ and $0.003 \leq k[h/{\rm Mpc}] \leq 15$.
    \label{f:dnu_class_comparison_Mnu150_DO}
  }
\end{figure}

\begin{table}
  \caption{
    $\nu\Lambda$CDM models used for code tests.  Our convention is to label
    each model by the series, the sum of neutrino masses in meV, and the
    mass ordering.  Thus, F150NO has $\Mnu=150$~meV, normally ordered, with
    the other parameters corresponding to series F below.  Series
    E corresponds to {\protect \cite{Euclid:2022qde}};
    F to {\protect \cite{Upadhye:2024ypg}}; and
    Q to {\protect \cite{Villaescusa-Navarro:2019bje}}.
    \label{t:nLCDM_models}
  }
  \renewcommand{\tabcolsep}{2mm}
  \begin{footnotesize}
    \begin{tabular}{c|ccccc}
      series & $\Omo$ & $\Obo$ & normalization & $h$ & $\ns$
      \\
      \hline
      E & $0.3190$ & $0.0490$ & $\As=2.215\times 10^{-9}$ & $0.67$ & $0.9619$\\
      F & $0.3316$ & $0.0490$ & $\As=2.2\times 10^{-9}$ & $0.6766$ & $0.9665$\\
      Q & $0.3175$ & $0.0490$ & $\sigma_8=0.834$ & $0.6711$ & $0.9624$\\
    \end{tabular}
  \end{footnotesize}
\end{table}

Since our starting point, \EQ{e:d0_a_k_u_formal}, is a neutrino population linearly coupled to a predetermined $\dm(a,k)$, the best method of testing \fastnuf{} is to compute $\dm$ and $\dnu$ using an existing linear solver such as \classcode{}, apply \EQ{e:d0_a_k_u_fastnf} to $\dm$, and compare the result to $\dnu$.  Figure~\ref{f:dnu_class_comparison_Mnu150_DO} makes this comparison at a range of $a$ and $k$ for the F150DO $\nu\Lambda$CDM model of Table~\ref{t:nLCDM_models}.  We evaluate $\dm(a,k)$  using \classcode{} at $a_j=0.01(j+1)$ for $0 \leq j \leq 99$.  Our \fastnuf{} computation of $\dnu$ integrates \EQ{e:def_dnu} using $50$-point Gauss-Laguerre quadrature, as implemented in \cite{Upadhye:2024ypg}, with flow velocities $u_\alpha = q_\alpha \Tnuo / m_\nu$ and $q_\alpha$ the quadrature points.  At $a=1$, the fractional difference remains below a percent at all wave numbers under $k=18~h/$Mpc, and below a tenth of a percent for $ 0.005~h/{\rm Mpc} \leq k \leq 4~h/{\rm Mpc}$.

Considering all scale factors shown in Fig.~\ref{f:dnu_class_comparison_Mnu150_DO}, there are two regimes where the error exceeds $1\%$, at low and high wave numbers.  At low $k$, \fastnuf{} errors are dominated by our assumption of non-relativistic neutrinos.  These $m_\nu = 50$~meV neutrinos go non-relativistic at $\anu \approx 0.01$.  Using an EdS growth approximation, we estimate the magnitude of this error as $\anu / a$, and we expect it to peak at scales $k \lesssim \knr \approx 0.004~h/$Mpc, in agreement with the figure.

Meanwhile, at high $k$, the difference becomes both larger and noisier with rising $k$, due to the combination of two numerical errors.  Neutrino clustering at small scales is dominated by low-velocity flows, meaning that increasing the Gauss-Laguerre quadrature order should reduce this error.  However, a second source of error is the finite numerical precison in our computations of the sine and cosine integrals.  Unfortunately, increasing the quadrature order to mitigate the former error exacerbates the latter.  Our choice of $50$-point quadrature provides sub-percent-level errors at all scale factors shown for $0.003~h/{\rm Mpc} \leq k \leq 15~h/{\rm Mpc}$.

We can test \fastnuf{} in a self-contained manner if we restrict ourselves to computing the ratio $\dnu / \dm$, since the normalization of these density contrasts depends upon early-universe physics for which we must rely upon another linear solver.  Since we do not know $\dm(a,k)$ from \fastnuf{} alone, even up to a constant factor, our solution must be iterative.  We begin by approximating $\dm$ as the matter growth factor $D_{\rm m}(a,k)$ estimated using \EQ{e:D_a_k_u},%
\begin{equation}
  D_{\rm m}(a,k)
  \approx
  \frac{\Ocbo}{\Omo} D(a,k,0)
  + \sum_\alpha \frac{\Onao}{\Omo} D(a,k,u_\alpha),
  \label{e:Dm_approx}
\end{equation}
where the flow density fractions $\Onao$ are found from the quadrature weights as in \cite{Upadhye:2024ypg}.  Substituting this for $\dm$ in \EQ{e:def_g_s_k}, we obtain $\delta(a,k,u)$ from \EQS{e:d0_a_k_u_fastnf}{e:Sigc} for $u=0$ as well as the quadrature flows $u_\alpha$.  This is our first iteration.  We may find the total neutrino density contrast as
\begin{equation}
  \dnu(a,k)
  =
  \sum_\alpha \frac{\Onao}{\Ono} \delta(a,k,u_\alpha),
\end{equation}
and $\dm$ by substituting $\delta(a,k,u)$ for $D(a,k,u)$ in \EQ{e:Dm_approx}.  We may iterate again using this $\dm$.  However, we find for the $\Mnu=150$~meV (DO) model considered above that a single iteration achieves sub-percent-level accuracy in $\dnu/\dm$ at $a=1$, compared with \classcode{}, for all $k \leq 20~h/$Mpc.

\begin{figure}
  \hskip-2mm\includegraphics[width=90mm]{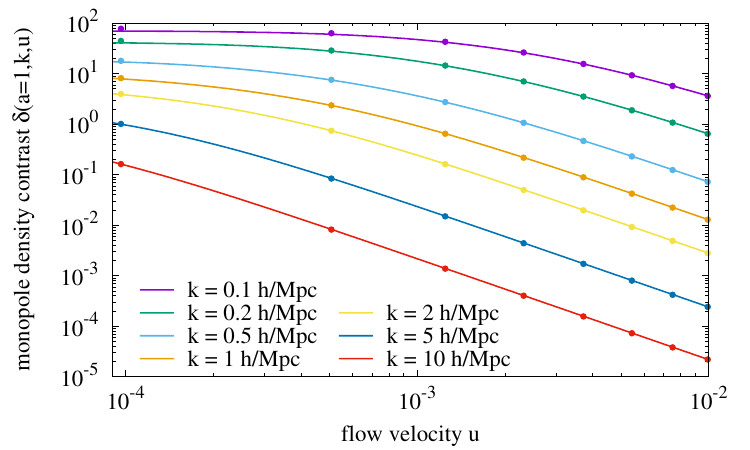}%
  \caption{
    Accuracy of \fastnuf{} for individual neutrino flows with a non-linear $\dcb$ source.
    \fastnuf{} flows (lines) closely match flows computed using the \muflr{} linear response code of {\protect\cite{Chen:2020bdf}} (points), for a
    $\nu\Lambda$CDM model with $\Mnu=150$~meV (NO) at $z=0$.  \fastnuf{} is as accurate as \muflr{} but  much faster, allowing us to compute $\delta(a,k,u)$ for many flows.
    \label{f:dnu_u_vs_MuFLR}
  }
\end{figure}

Neutrino linear response is equally applicable to a non-linear $\dm$.  Figure~\ref{f:dnu_u_vs_MuFLR} compares \fastnuf{} to the \muflr{} code of \cite{Chen:2020bdf}, which uses Time-RG perturbation theory to compute the non-linear $\dcb$, and then \EQS{e:eom_delta}{e:eom_theta} for the density contrasts of individual flows.  This test anticipates our application to emulation in Sec.~\ref{sec:emu}.  We find that fractional differences are $<10\%$ across three orders of magnitude in $k$ and two orders of magnitude in $u$.  Larger errors are found for $u \gtrsim 0.1$, where our assumption of non-relativistic neutrinos breaks down.

\section{Non-linear enhancement ratio}
\label{sec:Rnl}

\subsection{Properties of enhancement ratio}
\label{sub:Rnl:prop}

We proceed to study the non-linear enhancement ratio
\begin{equation}
  \Rnu(a,k,u) := \delta^{\rm (NL)}(a,k,u) ~/~ \delta^{\rm (LR)}(a,k,u),
  \label{e:Rnu_a_k_u}
\end{equation}
in perturbation theory, where NL and LR represent, respectively, the non-linear \fftm{} and linear-response \fastnuf{} computations.  It has a relatively low dynamic range: for the slowest $80\%$ of the neutrinos' Fermi-Dirac distribution, in each of the 101 cosmological models in Tables~C2-C4 of \cite{Moran:2022iwe}, in the ranges $0.25 \leq a \leq 1$ and $10^{-3} \leq k[h/{\rm Mpc}] \leq 1$, we find that $0.726 \leq \Rnu(a,k,u) \leq 2.455$, making it fairly simple to estimate, interpolate, or emulate.   If $\Rnu$ can be approximated as a function of $u$, then multiplying it by the \fastnuf{} density contrast will yield an approximation to the non-linear density contrast.  After studying $\Rnu$ in this Section, we will emulate it in the next.

\begin{figure}
  \hskip-3mm\includegraphics[width=92mm]{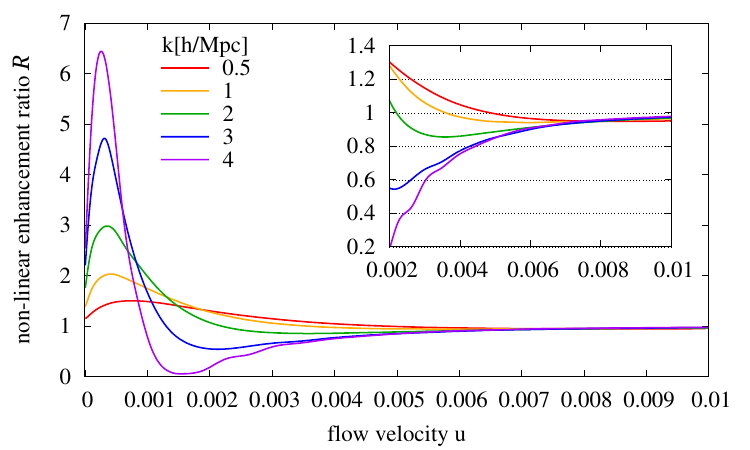}%
  \caption{
    Broad features of the non-linear enhancement ratio    $\Rnu(a,k,u)$ of \EQ{e:Rnu_a_k_u} at $a=1$, for the E600DO model of Table~\ref{t:nLCDM_models}.  For $k\leq 1~h/$Mpc, it has a low dynamic range, making interpolation accurate.
    The inset shows the high-$u$ region in greater detail, confirming that $\Rnu<1$ at high $u$.
    \label{f:Rc_wide}
  }
\end{figure}

Figure~\ref{f:Rc_wide} shows $\Rnu$ for a model with $\Mnu = 600$~meV.  Over a range of wave numbers, its behavior is qualitatively similar as $u$ is increased. The enhancement ratio rises with $u$ to a peak, then falls to a trough below unity before asymptotically approaching it at large $u$.  At larger $k$, both the peak and the trough grow and shift to lower $u$.  Note that the apparent oscillations in $\Rnu$ at high $u$, most evident at high $k$, are artifacts of the non-linear calculation of \fftm{}, which includes non-linear corrections for higher $u$ only at later times in order to avoid numerical instabilities.

The dip in $\Rnu$ below one, evident in the inset of Fig.~\ref{f:Rc_wide}, is the generalization to neutrinos of the well-known result from Standard Perturbation Theory that the non-linear correction is negative at low wave numbers.  Often attributed to power flowing from linear to non-linear scales, this is caused by the dominance at low $k$ of the $P^{(1,3)}(k)<0$ power spectrum correction, which gives way to the $P^{(2,2)}(k)>0$ term at larger $k$.  We verify this by running \fftm{} with only the $P^{(1,3)}$-like terms\footnote{The subtlety is that the FFTLog-based calculations of \cite{McEwen:2016fjn,Fang:2016wcf,Schmittfull:2016jsw,Upadhye:2017hdl,Chen:2022cgw} mix $P^{(1,3)}$-like and $P^{(2,2)}$-like terms in order to cancel divergences.  We restore the $P^{(1,3)}$-like terms.} and find $\Rnu<1$ at low $u$.

\subsection{Slow neutrinos}
\label{sub:Rnl:slow}

\begin{figure}
  \hskip-3mm\includegraphics[width=92mm]{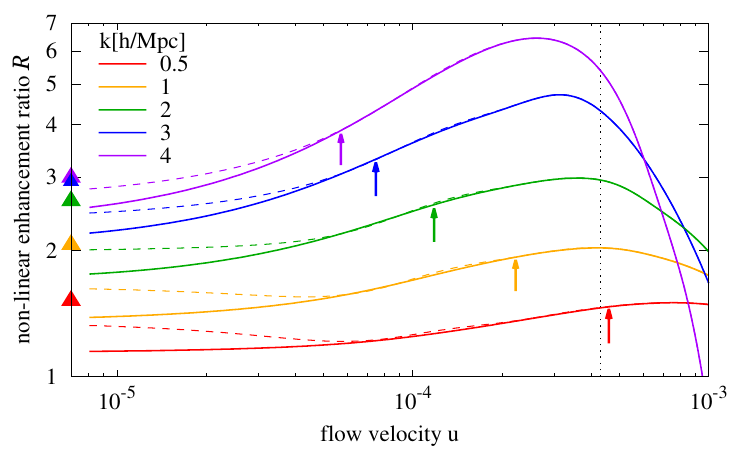}%
  \caption{
    Breakdown of the \fftm{} closure approximation of {\protect \cite{Chen:2022cgw}}, which assumes that neutrino non-linearity is small below $k=\kfs(a,u)$ (identified by vertical arrows).  $\Rnu(a,k,u)$ for the E600DO model of Table~\ref{t:nLCDM_models} (solid lines), precisely corresponding to that of Fig.~\ref{f:Rc_wide}, disagrees with the $u=0$ Time-RG points (triangles along the vertical axis).  An alternative closure approximation (dashed lines; see text) improves agreement with Time-RG at low $u$ without modifying $\Rnu$ to the right of the arrows, showing this region to be insensitive to our chosen approximation.  This motivates a simple interpolation between the Time-RG point and the $k \geq \kfs(a,u)$ region in order to avoid underpredicting $\Rnu$.  For reference, the region to the left of the vertical dotted line represents $1\%$ of the neutrino number density.
    \label{f:Rc_slow}
  }
\end{figure}

Next, we consider the slowest neutrinos, those with the smallest flow velocities.  Figure~\ref{f:Rc_slow} zooms in on the low-$u$ region of Fig.~\ref{f:Rc_wide}. Velocities below $0.00043$, shown by a vertical dotted line, correspond to the slowest $1\%$ of neutrinos for this model.  Although \fftm{} is derived from Time-RG perturbation theory, we see immediately that it does not reproduce the Time-RG $\Rnu$, identified by triangles in the figure, in the low-$u$ limit.

We believe that this is due to a difference between the closure approximations made in Time-RG and \fftm{}.  In order to approximate the linear evolution of the bispectrum integrals as depending only upon the bispectrum integrals themselves, \cite{Chen:2022cgw} neglected the wave-number-dependence of the linear evolution matrix, which in ordinary Time-RG is
\begin{equation}
  \Xi_{ab}(k) =
  \left[ \begin{array}{cc}
      0 & -1 \\ \frac{k^2 \Phi}{\Hc^2 \delta} & ~~1+\frac{d\ln \Hc}{d \ln a}
      \end{array}  \right].
\end{equation}
If Time-RG is applied only to cold matter, then the lower-left component is simply $-\tfrac{3}{2}$ according to Poisson's equation; the $k$-dependence vanishes.  When neutrinos are included as linear sources, as in \cite{Lesgourgues:2009am}, it declines from $-\tfrac{3}{2}$ at small $k$ to $-\tfrac{3}{2}(1-\fnu)$ at large $k$, a variation which \cite{Upadhye:2015lia} found to be negligible for $\Ono h^2 \leq 0.01$.

However, when Time-RG is extended to neutrinos, $k^2 \Phi / (\Hc^2 \delta(a,k,u))$ is highly $k$-dependent, and $\Xi_{ab}(k)$ contains additional $k$-dependent free-streaming terms.  Closure requires us to neglect the dependence of $\Xi_{ab}(k')$ upon the integration wave number $k'$.  \cite{Chen:2022cgw} reasoned that, since non-linear corrections to neutrino clustering are important only well below the free-streaming scale, neutrinos will freely stream past gravitational potentials, motivating them to neglect both the $\Phi$ term and the free-streaming terms in the evolution of bispectrum integrals.

Non-linear corrections at wave number $k$ will approximately be in this free-streaming regime for $k > \kfs(a,u)$, that is, for $u$ larger than $u_{\rm TRG} := \Hco k^{-1} \sqrt{\tfrac{3}{2}\Omo a}$.  Vertical arrows in Fig.~\ref{f:Rc_slow} identify $u_{\rm TRG}$ for each curve.  Evidently, for $k \gtrsim 0.5~h/$Mpc, where non-linear corrections to neutrino clustering are important, this free-streaming closure approximation is valid for $\gtrsim 99\%$ of the neutrinos.  However, for $u \ll u_{\rm TRG}$, it will underestimate non-linear growth, consistent with the solid \fftm{} curves underpredicting the Time-RG points in the figure.

Although closure requires that $\Xi_{ab}(k')$ be independent of the integration wave number $k'$, we may artificially impose a dependence upon the external wave number $k$.  Motivated by the fitting function of \EQ{e:supp_nu}, we make the crude replacement $\tfrac{(k')^2 \Phi(k')}{\Hc^2 \delta(k',u)} \rightarrow \tfrac{k^2 \Phi(k)}{\Hc^2 \delta(k,u)} (1 + k/\knr)^{-2}$ in the lower-left entry of $\Xi_{ab}(k')$.  By construction, this will approach the Time-RG value of $-\tfrac{3}{2}$ at low $k$ and suppress gravitational clustering at high $k$.  Replacing the \fftm{} closure approximation by this one, we find the dashed curves in Fig.~\ref{f:Rc_slow}.  As expected, they make $\Rnu$ significantly closer to its Time-RG value at low $u$ while having a negligible effect for $u \geq u_{\rm TRG}$.

Encouraged by this success, one might consider a fitting function for the closure approximation.  The difficulty is that we have no calculation against which to calibrate it.  \cite{Chen:2022cgw} argued that, in the absence of a closure approximation, they would have to track the full functional dependence of the bispectrum upon three wave numbers and two angles, resulting in a coupled system of over a billion equations for each flow, a computationally prohibitive task.

Instead, we regard the close agreement between the two closure approximations in the $u \gtrsim u_{\rm TRG}$ region as confirmation of the accuracy of \fftm{} there.  Combined with Time-RG as a calculation of $\Rnu$ at $u=0$, we can see that the qualitative rise-fall-rise behavior of $\Rnu$ vs.~$u$ in Fig.~\ref{f:Rc_wide} is accurate at least for $k \gtrsim 2~h/$Mpc, since the peaks of $\Rnu$ in Fig.~\ref{f:Rc_slow} occur at $u > u_{\rm TRG}$ and have $\Rnu$ greater than the corresponding Time-RG points.  For $k \lesssim 1~h/$Mpc, the closeness between the $\Rnu$ values at $u=0$ and at the peak is consistent with a flat low-$u$ $\Rnu$, as well as one which falls at low $u$ before rising again.

\subsection{Approximation of enhancement ratio}
\label{sub:Rnl:approx}

\begin{figure}
  \hskip-3mm\includegraphics[width=92mm]{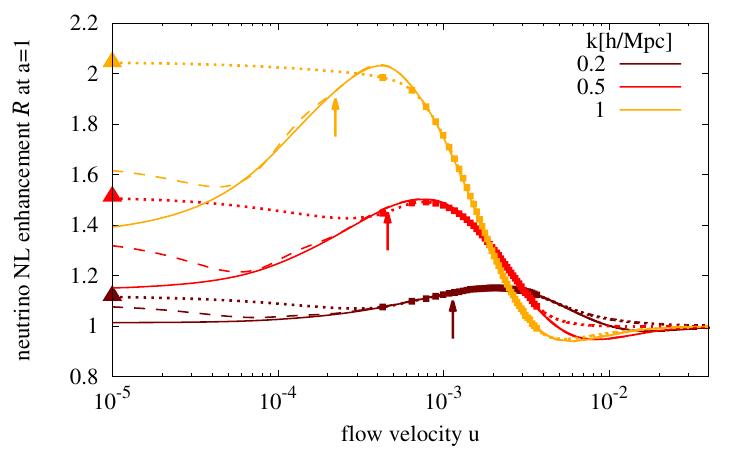}%
  \caption{
    Interpolation and extrapolation of the non-linear enhancement ratio $\Rnu(a,k,u)$ for the E600DO model of Table~\ref{t:nLCDM_models}. At low $u$, interpolation (dotted lines) between the Time-RG values (large triangular points) and the \cosmicenu{} training data (small square points) avoids the underpredictions of $\Rnu$ evident in the  solid and dashed lines, corresponding to the two closure approximations of Fig.~\ref{f:Rc_slow}. The training data approximately cover the free-streaming region $k \geq \kfs(a,u)$, to the right of the vertical arrows. Meanwhile, at high $u$, extrapolation (dotted lines) using the training data matches direct calculation of $\Rnu$ (solid lines) to a few percent.
    \label{f:Rc_int_ext}
  }
\end{figure}

Although we do not have a rigorous calculation of $\Rnu(a,k,u)$ between  $u=0$ and $u_{\rm TRG}$, we see for $k \leq 1~h/$Mpc that $\Rnu$ agrees to $<10\%$ at the two velocities.  Thus, a simple interpolation between the two is a reasonable approximation.  In practice, the lowest non-zero $m_\nu u$ available in the \cosmicenu{} training data, $0.086$~meV, corresponds to the vertical dotted line in Fig.~\ref{f:Rc_slow}, $u \approx 0.00043$ for this model, while the next two are $0.129$~meV and $0.157$~meV.  We approximate $\Rnu$ using a cubic polynomial interpolating $u=0$ and the first three \cosmicenu{} velocities.  Figure~\ref{f:Rc_int_ext} illustrates this interpolation.  Points show the \cosmicenu{} training points, triangles the $u=0$ Time-RG $\Rnu$, and dotted lines our interpolation between them at low $u$.

For $k \ll 0.5~h/$Mpc, one may worry that $u=0.00043$ is much less than $u_{\rm TRG}$, making this point unreliable.  This is mitigated by two factors: the smallness of $\Rnu-1$ at low $k$, since non-linearity is weak there; and the relative insensitivity of the low-$k$ total neutrino power spectrum to the slowest neutrinos, since faster ones still contribute significantly to the power there.  The first is demonstrated by the $k=0.2~h/$Mpc curve in Fig.~\ref{f:Rc_int_ext}, for which $|\Rnu-1|\leq 0.15$.

The largest \cosmicenu{} training momentum is $1.437$~meV, corresponding to $u=0.0072$ for the E600DO model.  We will see in the next Section that the fastest flows are the most affected by numerical errors, since they cluster much more weakly than the slower ones.  Thus we will discard the fastest ten, leaving $m_\nu u = 0.733$~meV the fastest momentum considered here, corresponding to $u=0.0037$ for this model.  In either case, we may see from the inset to Fig.~\ref{f:Rc_wide} that $\Rnu$ differs from unity by several percent even for $u$ greater than this.

We extrapolate to higher $u$ using the function
\begin{equation}
  \Rnu_{\rm hi} = 1 + \frac{A_2}{u^2} + \frac{A_3}{u^3},
\end{equation}
which approaches unity at large $u$ by design.  This is fit to the \cosmicenu{} training data at two points: the highest-velocity point considered, with a momentum $0.733$~meV; and either the velocity of the first minimum $\Rnu<1$, or the second-highest velocity.  This extrapolation is shown by dotted lines in Fig.~\ref{f:Rc_int_ext}.

Note one source of error apparent in the figure.  At high $u$, our extrapolated approximation correctly fits the dip below unity, and the subsequent rise, only if the \cosmicenu{} training data reach $\Rnu<1$.  For $k \lesssim 0.5~h/$Mpc, where they do not, a $\approx 5\%$ error is evident in the figure at high $u$.  A more accurate study of fast neutrinos would, therefore, require further calculation.

\section{Momentum-space emulation}
\label{sec:emu}

\subsection{Motivation: small-scale neutrino power}
\label{sub:emu:motivation}

\begin{figure}
  \includegraphics[width=87mm]{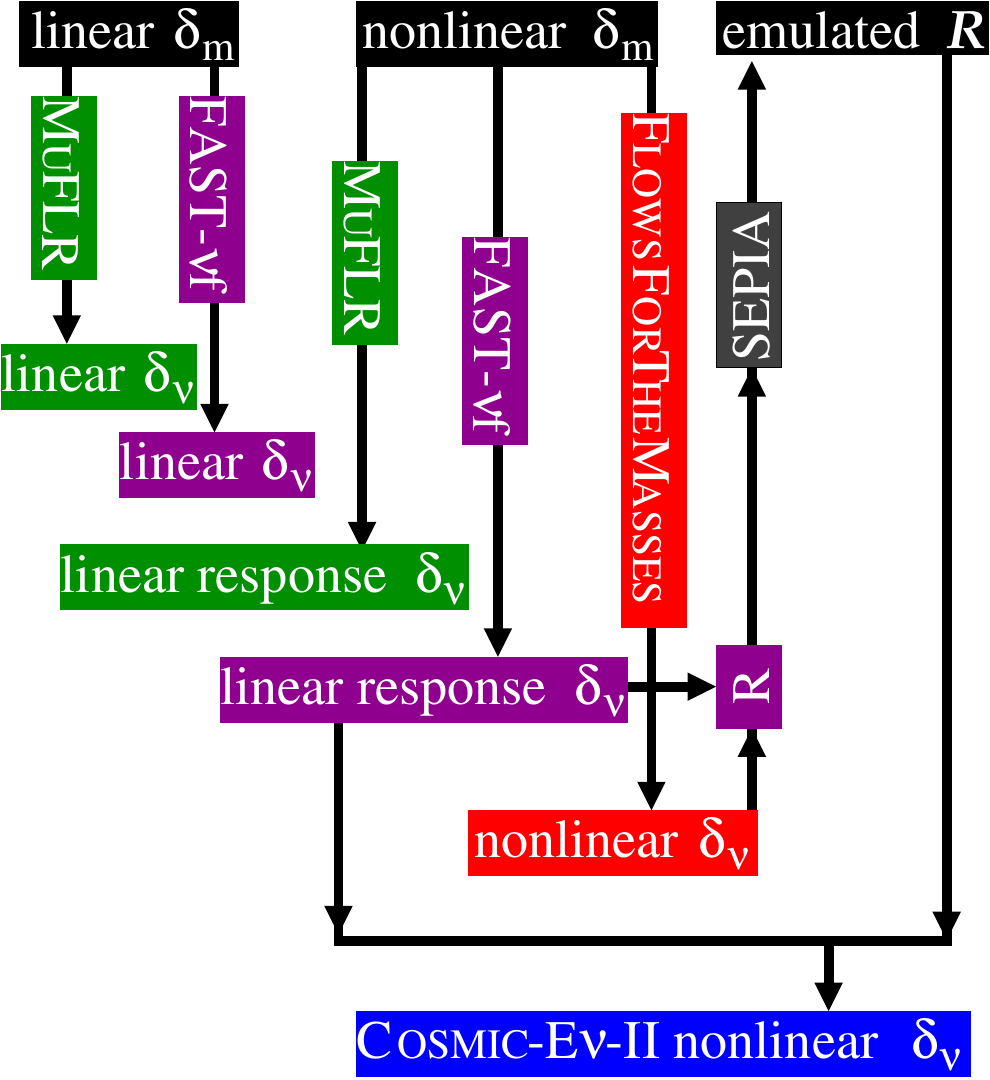}%
  \caption{
    A guide to the different neutrino density contrasts considered here.
    For a problem size of $\sim 10$ flows and $\sim 100$ wave numbers,
    running on a typical desktop machine, \muflr{} of
    {\protect \cite{Chen:2020bdf}} applied to a linear (non-linear) $\dm$
    takes minutes to compute the linear (linear response) $\dnu$, while
    \fastnuf{} of Sec.~\ref{sec:fnf} takes only milliseconds.  \fftm{} requires
    hours but recovers the non-linear $\dnu$.  Dividing this non-linear $\dnu$
    by the \fastnuf{} linear response $\dnu$, then using {\sc sepia} of
    {\protect \cite{SEPIA}} to emulate the resulting
    $\Rnu$ (\enuii{}), we can recover the non-linear $\dnu$ in tens of
    milliseconds.  The \fastnuf{} linear $\dnu$ was tested in
    Fig.~\ref{f:dnu_class_comparison_Mnu150_DO}, and the \fastnuf{} linear
    response $\dnu$ in Fig.~\ref{f:dnu_u_vs_MuFLR}, while the non-linear $\dnu$
    resulting from \enuii{} will be compared with N-body simulations in
    Sec.~\ref{sub:emu:enuii}.
    \label{f:enuii_flowchart}
  }
\end{figure}

Our main result of this Section will be to use the \fastnuf{} linear response method of Sec.~\ref{sec:fnf} to improve the momentum resolution of the \cosmicenu{} emulator of \cite{Upadhye:2023bgx}.  Figure~\ref{f:enuii_flowchart}  summarizes the linear, linear response, and non-linear methods of computing the neutrino density contrast $\dnu$; their relative computational expenses; and the combination of \fastnuf{} linear response with the emulated \fftm{} $\Rnu(a,k,u)$ in \enuii{}.

Slow flows are important, as shown by \cite{Upadhye:2024ypg}, because they dominate small-scale neutrino clustering, particularly for low $\Mnu$.  Specifically,  for model E150DO, a greater low-momentum sampling reduced the error in the neutrino power spectrum from $21\%$ to $5.4\%$ at $k=0.4~h/$Mpc, and $50\%$ to $22\%$ at $k=1~h/$Mpc, relative to the N-body simulations of~\cite{Euclid:2022qde}.

However, these low-$u$ flows exacerbate the numerical instabilities of the \fftm{} perturbation theory, which are particularly severe at high $\Mnu$.  Despite its progress in mitigating these instabilities, \cite{Upadhye:2024ypg} was unable to apply this low-$u$ sampling to $932$~meV, the upper mass limit of \cosmicenu{}, and required some adjustment even at $600$~meV.  Interpolating lower-$u$ flows using Time-RG provides this sampling without the attendant instabilities.

Our strategy, then, is to emulate $\Rnu(a,k,u)$ using the existing training data, interpolate to low $u$, and extrapolate to high $u$, as discussed in Sec.~\ref{sec:Rnl}.  Multiplying this by $\delta^{\rm (LR)}(a,k,u)$ recovers $\delta^{\rm (NL)}(a,k,u)$.  Until now, the chief difficulty with this approach is the lack of a fast momentum-space linear response perturbation theory.  We will show in this Section that \fastnuf{} is well-suited for this purpose.  It significantly improves emulator accuracy for slow neutrinos, hence small scales and $\Mnu$.  Furthermore, accurate interpolation allows us to predict the clustering of any range of neutrino momenta, not just the deciles emulated by \cosmicenu{}.  This could be useful, for example, in a hybrid N-body simulation code in which only the slowest neutrinos were tracked using simulation particles.

\begin{table*}
  \caption{
    Time steps used in our \fastnuf{} application to $\dnu$ emulation in
    \enuii{}.  The earliest, $j=0$, uses linear perturbation theory.  The next
    twenty-four, $1 \leq j \leq 24$, use the interpolated growth approximation
    of Sec.~\ref{sub:emu:dm}.  The next two, $j$ of $25$ and $26$, use the
    Time-RG perturbation theory for $\dcb(a,k)$, and the remaining steps use
    the \MTiv{} emulator. At all time steps $j \geq 25$, $\dnu(a,k)$ is computed
    using the \fftm{} perturbation theory.  Steps $j=0$ and $25$-$51$ are drawn
    from the training set of \cosmicenu{}.
    \label{t:time_steps_fastnuf}
  }
  \renewcommand{\tabcolsep}{2mm}
  \begin{footnotesize}
    \begin{tabular}{c|ccccccccccccccc}
      $j$ & $0$ & $1$-$2$4 & $25$ & $26$ & $27$
      & $28$ & $29$ & $30$ & $31$ & $32$
      & $33$ & $34$ & $35$ & $36$ & $37$
      \\
      $z_j$ & $200$ & $\tfrac{100}{j}-1$ & $3.04$ & $2.478$ & $2.02$
      & $1.799$ & $1.61$ & $1.376$ & $1.209$ & $1.006$
      & $0.779$ & $0.736$ & $0.695$ & $0.656$ & $0.6431$
      \\
      \hline
      $j$ & $38$ & $39$ & $40$ & $41$ & $42$
      & $43$ & $44$ & $45$ & $46$ & $47$
      & $48$ & $49$ & $50$ & $51$ & {}
      \\
      $z_j$ & $0.618$ & $0.578$ & $0.539$ & $0.502$ & $0.471$
      & $0.434$ & $0.402$ & $0.364$ & $0.304$ & $0.242$
      & $0.212$ & $0.154$ & $0.101$ & $0$ & {}
    \end{tabular}
  \end{footnotesize}
\end{table*}

The starting point is the training set of the \cosmicenu{} emulator, which uses the same $101$ massive-neutrino models used to train the \MTiv{} emulator of \cite{Moran:2022iwe}.  Its cosmological parameters are listed in Tables~C2-C4 of that reference.  For each, it finds density contrasts for the cb fluid as well as fifty neutrino flows, each one representing $2\%$ of the neutrino density, under the assumption of DO masses.  It provides this information at $78$ logarithmically-spaced wave numbers between $10^{-3}~h/$Mpc and $1~h/$Mpc, at $z=200$ as well as the final $27$ redshifts listed in Table~\ref{t:time_steps_fastnuf}, that is, $25 \leq j \leq 51$.

\subsection{Matter density contrast}
\label{sub:emu:dm}

\begin{figure}
  \hskip-3mm\includegraphics[width=91mm]{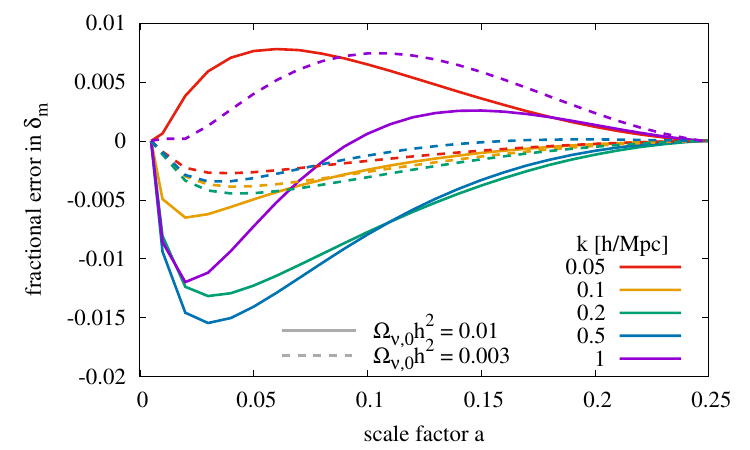}%
  \caption{
    The $\dm(a,k)$ interpolation of \EQ{e:dm_growth_expansion} is accurate
    to $\leq 1.5\%$, over a large range of $\Ono h^2$, compared to the
    corresponding Time-RG $\dm(a,k)$.
    The two $\nu\Lambda$CDM models shown have $\Omo h^2=0.15$, $\Obo h^2=0.022$,
    $\sigma_8=0.85$, $\ns=0.97$, $h=0.67$, and $\Ono h^2$ of $0.01$
    (solid lines) or $0.003$ (dashed lines).
    \label{f:dm_growth_expansion}
  }
\end{figure}

Our computation of $\dm(a,k)$ for DO neutrinos is straightforward given the existing \cosmicenu{} training set.  The cb density contrast is taken from the \MTiv{} emulator at $z \leq 2.02$, Time-RG perturbation theory at $2.02 < z \leq 3.04$, and linear perturbation theory at $z=200$.  To this we add the non-linear \fftm{} neutrino density contrast, with $\dcb$ and $\dnu$ weighted by $\Ocbo/\Omo$ and $\Ono/\Omo$, respectively.  In practice, switching from the linear $\dnu$ used in \MTiv{} to the non-linear one makes only a sub-percent-level difference, as shown by \cite{Upadhye:2023bgx}.

The emulation procedure itself was described thoroughly in \cite{Heitmann:2009cu,Lawrence:2009uk,Lawrence:2017ost,Moran:2022iwe}.  Its specific implementation in our emulator, using the {\sc sepia} software of \cite{SEPIA}, was documented in \cite{Upadhye:2023bgx}.  As we make no further modifications to the emulation procedure, we refer interested readers to those publications for details.

The \fastnuf{} neutrino density contrast of \EQ{e:d0_a_k_u_fastnf}, beginning with the time-integration of \EQ{e:d0_a_k_u_formal}, requires that $\dm(a,k)$ be sampled finely enough that $g(s,k)$ may be interpolated as a function of $s$.  However, the high-redshift behavior $s \propto a^{-1/2}$ makes it suboptimal for interpolating $\dm$, hence $g$, between the training points at $z=200$ and $3.04$. Our approach is to fit $\dm$ at these two redshifts, as well as $z=2.478$, using a third-order growth expansion,%
\begin{eqnarray}
  \dm(a,k) &=& \sum_{m=1}^3 c_m(k) D_{\rm m}(a,k)^m
  \label{e:dm_growth_expansion}
  \\
  D_{\rm m}(a,k) &=&
  \left[\frac{\Ocbo}{\Omo} + \frac{\Ono}{\Omo}\xi(a,k,c_\nu)\right] D(a,k,0)
\end{eqnarray}
with $D(a,k,0)$ found by setting $u=0$ in \EQ{e:D_a_k_u}.  Standard Perturbation Theory at $1$-loop-order is itself a third-order expansion in the linear $\dm$, and \cite{Saito:2009ah} derives the time-dependence of the density contrast in the presence of massive neutrinos, showing that it reduces to the form \EQ{e:dm_growth_expansion} in the EdS limit.  Thus this is an adequate approximation to the mildly non-linear growth at $z \gtrsim 3$ while guaranteeing linear growth at small $a$.  In order to accommodate the $s$ interpolation required for \fastnuf{}, we use this growth expansion to add points at scale factors $a_j = 0.01 j$ for integers $1 \leq j \leq 24$.  Figure~\ref{f:dm_growth_expansion} confirms its accuracy at precisely these scale factors for two $\nu\Lambda$CDM models with $\Ono h^2$ of $0.01$ and $0.003$.

Since our training set uses the degenerate-mass neutrino calculations of \cite{Moran:2022iwe,Chen:2022cgw}, applying it to NO or IO neutrinos is not strictly correct.  Thus we next derive correction factors approximating the linear $\dmNO(a,k)/\dmDO(a,k)$ and $\dmIO(a,k)/\dmDO(a,k)$.  Our task is simplified by the fact that any neutrino corrections to $\dm$ are necessarily suppressed by their fraction $\fnu$, while NO and IO differ the most from DO precisely at the smallest $\fnu$.  As a result, the maximum error associated with using $\dmDO$ for NO or IO neutrinos is only $\approx 0.2\%$, and peaks at low $\Mnu$.  We will show that this error may be reduced by an order of magnitude at late times and small scales, rendering it negligible.

We begin with the linear growth approximation of \EQ{e:D_a_k_u}.  Let $\Omega(u)$ be the flow-dependent neutrino density fraction, such that neutrinos with velocities between $u$ and $u+\Delta u$ make up a density fraction $[d\Omega(u)/du] \Delta u$.  Then the total-matter growth factor implied by \EQ{e:D_a_k_u} is $D_{\rm m}(a,k) = (D+2a_{\rm mr}/3) \Omo^{-1} \supp(a,k) [\Ocbo + \int du (d\Omega/du) \gsef(a,k,u)]$.  Only $\supp(a,k)$ and the factor in square brackets depend upon the neutrino mass ordering, the latter through $d\Omega/du$.  We consider each of these in turn.

First, consider $\supp(a,k)$.  From \EQEQ{e:def_Qnu_anu}{e:def_knr} we see that $\Qnu \approx \tfrac{3}{5}\fnu$ is approximately linear in the neutrino density fraction, hence each of the individual neutrino masses; $\anu$ is inversely proportional to $m_\nu$; and $\knr \propto \sqrt{m_\nu}$.  Let $\mnus$ be the mass of a neutrino of species $s$, with $\sum_s \mnus = \Mnu$.  Thus we replace \EQ{e:supp_nu} by a sum of the individual-species contributions:
\begin{eqnarray}
  \supp(a,k)
  &=&
  1 -
  \sum_s \frac{\Qnus \log(a/\anu(\mnus)) k^2
  }{\beta_0 \knr(\mnus)^2 + \beta_1 \knr(\mnus)\, k + k^2}
  \\
  \Qnus
  &=&
  \frac{5}{4} \left(1 - \sqrt{1 - \frac{24}{25}\frac{\mnus}{\Mnu}\fnu}\right)
  \approx \frac{3}{5} \frac{\mnus}{\Mnu} \fnu~.
\end{eqnarray}
For DO neutrinos, this matches \EQ{e:supp_nu} at small $\fnu$.

Next, consider $d\Omega/du$, which, for a single non-relativistic $\nu {\bar \nu}$ pair of mass $m_\nu$, was shown in \cite{Upadhye:2024ypg} to be%
\begin{equation}
  \frac{d\Omega(u)}{du}
  =
  \frac{m_\nu^4}{\pi^2{\bar \rho}_{\mathrm{crit},0}} u^2 f(m_\nu u).
\end{equation}
This expression is the normalized integrand of ${\bar T}^0_0$ from that reference, though note that our distribution function definition is $(2\pi)^3$ times theirs.
Here, ${\bar \rho}_{\mathrm{crit},0} = 3\Hco^2 / (8\pi G_{\rm N})$ is the critical density of the universe today.  For DO neutrinos this is simply tripled, with $f$ set to the Fermi-Dirac distribution, while the NO and IO cases may be treated using their respective effective distribution functions, \EQ{e:def_f_ehdm}, as described in Sec.~\ref{sub:bkg:ehdm}.

Thus our linear correction factor for $\dmDO(a,k)$ is
\begin{equation}
  \frac{\dm^{\rm (XO)}}{\dmDO}
  \approx
  \frac{\supp^{\rm (XO)}(a,k)}{\supp^{\rm (DO)}(a,k)}
  \,
  \frac{\Ocb + \int du \frac{d\Omega^{\rm (XO)}(u)}{du} \gsef(a,k,u)
  }{\Ocb + \int du \frac{d\Omega^{\rm (DO)}(u)}{du} \gsef(a,k,u)}
  \label{e:dmDO_correction_factor}
\end{equation}
where XO represents either NO or IO.  Beginning with our emulated $\dmDO(a,k)$, we multiply by \EQ{e:dmDO_correction_factor} to approximate $\dmNO(a,k)$ and $\dmIO(a,k)$.  Standard quadrature schemes allow us to represent the integral over $d\Omega/du$ using a finite number of flows.  We use Gauss-Laguerre quadrature, appropriate to exponentially-decaying $f(m_\nu u)$, as in \cite{Upadhye:2024ypg}.

\begin{figure}
  \hskip-4mm\includegraphics[width=93mm]{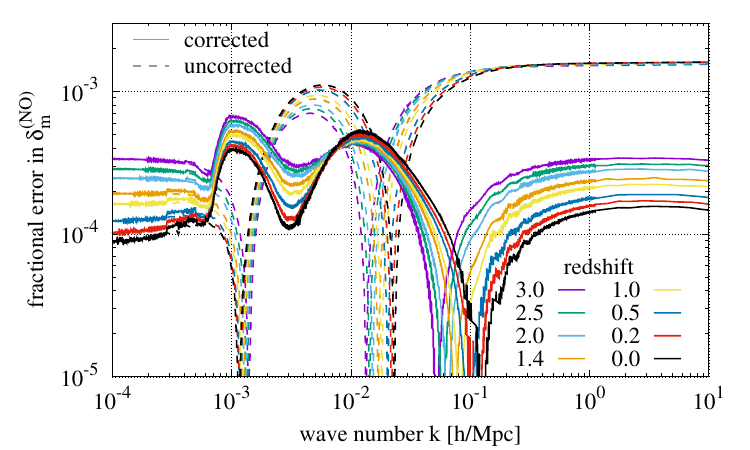}%

  \hskip-4mm\includegraphics[width=93mm]{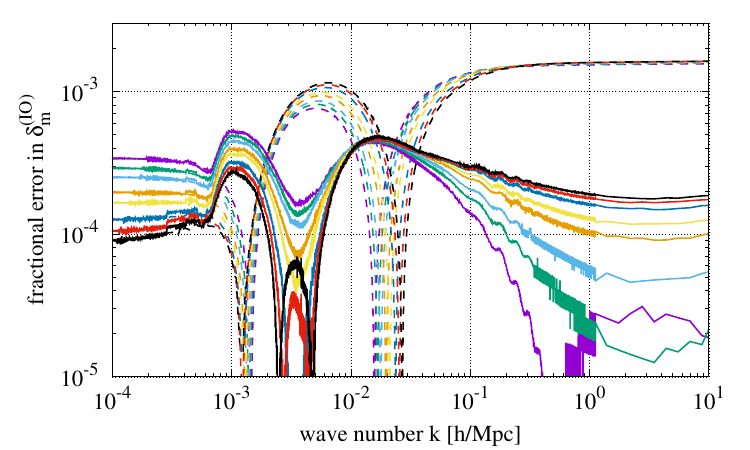}%
  \caption{
    Error reduction through the $\dm$ correction factor of
    \EQ{e:dmDO_correction_factor}, which allows $\dm$ computed or emulated for
    DO neutrinos to be applied to NO and IO neutrinos in \fastnuf{}.
    Minimum-$\Mnu$ $\dmNO(a,k)$ (top) and $\dmIO(a,k)$ (bottom) are
    approximated using either $\dmDO(a,k)$ alone (dashed) or $\dmDO(a,k)$ times
    \EQ{e:dmDO_correction_factor} (solid), and compared to the respective
    matter density contrasts computed using \classcode{}.  Applying this
    correction  reduces the $k\geq0.1~h/$Mpc error by a factor of
    at least five.
    \label{f:dmDO_correction_factor}
  }
\end{figure}

Figure~\ref{f:dmDO_correction_factor} shows our results at the minimum $\Mnu$ for each mass ordering.  Our correction reduces the small-scale errors, $k \geq 0.1~h/$Mpc, in both cases, by factors of five to ten.  Though the errors in using $\dmDO$ to approximate $\dmNO$ or $\dmIO$ are small, $\approx 0.16\%$ at small scales, these will be doubled in the power spectrum and further increased by late-time non-linear growth.  Thus we apply these corrections despite their smallness.  These errors fall off rapidly with $\Mnu$, and our correction continues to reduce them further.  For example, at $\Mnu=90$~meV, the maximum error across all $z$ and $k$ shown falls by a factor of $\approx 3$, from $0.052\%$ to $0.018\%$.

Since recent analyses such as \cite{DESI:2025ejh,SPT-3G:2025zuh} prefer masses below the minimum values of Sec.~\ref{sub:bkg:mass_order}, we define a prescription for reducing $\Mnu$ below these minima while ensuring the continuity and non-negativity of each individual mass.  In the NO case, for $\Mnu$ below $\sqrt{\Delta m_{31}^2} = 50.1$~meV, we fix $m_{\nu,1}$ and $m_{\nu,2}$ to zero while setting $m_{\nu,3}$ to $\Mnu$.  Between $50.1$~meV and $\sqrt{\Delta m_{31}^2} + \sqrt{\Delta m_{21}^2} = 58.8$~meV, we set $m_{\nu,1}$ to zero, neglect the smaller mass splitting, fix $m_{\nu,3}$ to $\sqrt{\Delta m_{31}^2}$, and choose $m_{\nu,2} = \Mnu - m_{\nu,3}$.

In the IO case, for $\Mnu$ below $2 \sqrt{-\Delta m_{32}^2 - \Delta m_{21}^2} = 98.2$~meV, we set $m_{\nu,3}$ to zero, and each of $m_{\nu,1}$ and $m_{\nu,2}$ to $\Mnu/2$.  Between $98.2$~meV and $\sqrt{-\Delta m_{32}^2} + \sqrt{-\Delta m_{32}^2 - \Delta m_{21}^2} = 98.9$~meV, we set $m_{\nu,3}$ to zero, $m_{\nu,1}$ to $\sqrt{-\Delta m_{32}^2 - \Delta m_{21}^2} = 49.1$~meV, and choose $m_{\nu,2} = \Mnu-m_{\nu,1}$.  For example, if $\Mnu=30$~meV, our procedure chooses a single massive $30$~meV neutrino and two massless ones for NO; two massive $15$~meV neutrinos and a massless one for IO; and three massive $10$~meV neutrinos for DO.

\subsection{Neutrino enhancement ratio}
\label{sub:emu:Rnu}

Emulator training sets typically divide the data to be emulated by a fast, accurate approximation to those data.  This reduction in dynamic range means that the emulator need only approximate a small correction to the approximation, rather than the whole functional form of the data.  The training set of \cosmicenu{} divided the momentum-space neutrino densities by linear approximations applying \EQS{e:gsef_a_k_u}{e:gsef_a_k_cnu} to the matter density of \cite{Eisenstein:1997jh,Eisenstein:1997ik}.  Since this approximation does not include non-linear growth of the cb fluid, it significantly underestimates neutrino overdensities at small scales, leading to a large dynamic range which degrades emulator accuracy. The \muflr{} code, allowing neutrino flow overdensities to respond linearly to the non-linear $\dm$, is more accurate; however, its running time of several minutes makes it too expensive for emulator applications.

The \fastnuf{} method of Sec.~\ref{sec:fnf} is a fast, accurate multi-fluid neutrino simulation method, making it applicable to momentum-space neutrino emulation.  Its consistency with \muflr{} for a broad range of scales and velocities was shown in Fig.~\ref{f:dnu_u_vs_MuFLR}.  For a given cosmological model, we begin with the $\dm$ emulated in Sec.~\ref{sub:emu:dm}.  Applying the \fastnuf{} method, we compute the linear response $\delta^{\rm (LR)}(a_j,k,u)$ for all time steps $a_j$ in Table~\ref{t:time_steps_fastnuf}, all emulated $k$ values, and all equal-density flow velocities $u_\beta$ in the \cosmicenu{} training set.

Dividing the \cosmicenu{} training data by this $\delta^{\rm (LR)}(a_j,k,u)$, we compute the non-linear enhancement ratio $\Rnu(a,k,u)$ of \EQ{e:Rnu_a_k_u}, the quantity which we emulate for $10^{-3} \leq k[h/{\rm Mpc}] \leq 1$.  Section~\ref{sec:Rnl} thoroughly studied the properties of $\Rnu$ and its approximation. Since we may interpolate the emulated $\Rnu$ in $u$, as described in that section,  we also compute $\delta^{\rm (LR)}(a_j,k,u)$ at additional velocities $u_\alpha$ with which we may improve the momentum sampling.  Unless otherwise stated, we choose these additional velocities using the $50$-point Gauss-Laguerre quadrature method.

\begin{figure}
  \hskip-4mm\includegraphics[width=92mm]{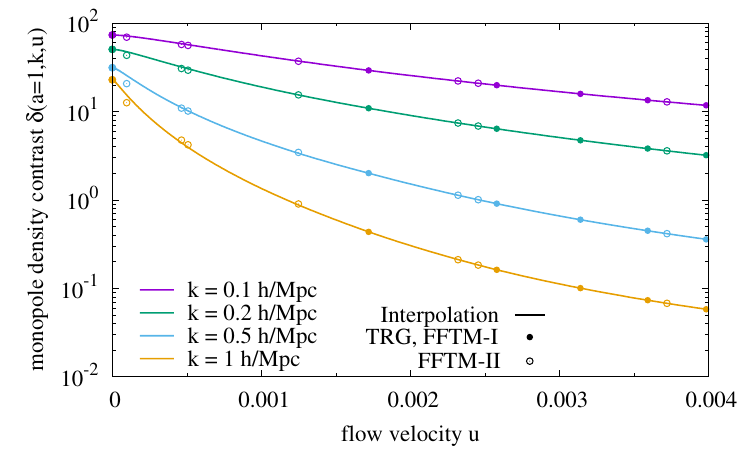 }%
  \caption{
    Interpolation (lines) of Time-RG and uniform-density \fftm{} flows
    (filled poins) used to predict the \fftm-II Gauss-Laguerre flows of
    \protect\cite{Upadhye:2024ypg} (open points) for a $\nu\Lambda$CDM model
    with $\Mnu=150$~meV (DO).  Interpolation closely matches the Gauss-Laguerre
    flows, aside from the $k \gtrsim 0.5~h/$Mpc, $u\ll 0.001$ region, where
    \fftm-II itself underpredicts $\dnu$, as discussed in Sec.~\ref{sec:Rnl}.
    \label{f:da0_interp_u}
  }
\end{figure}

Figure~\ref{f:da0_interp_u} illustrates the use of the interpolated $\Rnu(a,k,u)$, multiplied by the \fastnuf{} $\delta^{\rm (LR)}(a_j,k,u)$, to predict the non-linear density contrast over a range of flow velocities.  Interpolation allows us to improve the $u$ sampling of the non-linear $\delta(a,k,u)$ to match $50$-point or $10$-point Gauss-Laguerre velocities (open circles), up to low-$u$ discrepancies studied in Sec.~\ref{sec:Rnl}.  Note that $\delta(a,k,u)$ itself varies by orders of magnitude, making $\Rnu(a,k,u)$ a better choice for interpolation.

\begin{figure}
  \hskip-3mm\includegraphics[width=92mm]{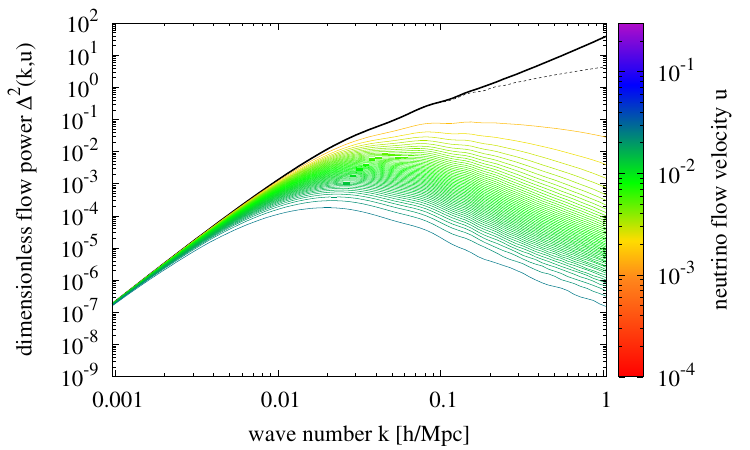}%

  \hskip-3mm\includegraphics[width=92mm]{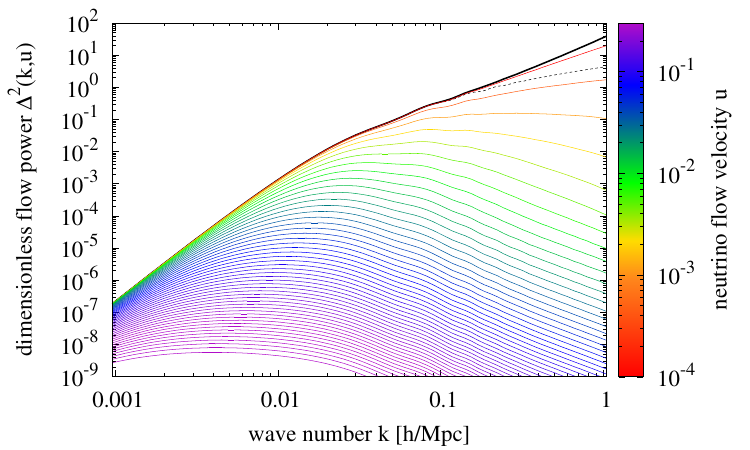}%
  \caption{
    Power spectra of fifty flows, corresponding to equal
    fractions of the neutrino number density (top) or Gauss-Laguerre
    quadrature points (bottom), for model 14 of {\protect \cite{Moran:2022iwe}},
    Tables~C2-C4, with $\Mnu\approx163$~meV, at $z=0$.  Solid and dashed black
    lines show Time-RG and linear cb power spectra, respectively.  Improved
    resolution at low $u$, evident in the lower plot, is particularly important
    for predicting neutrinos' small-scale clustering.
    \label{f:D2nu_unif_vs_GLQ}
  }
\end{figure}

The effect of this interpolation on the $u$-dependent clustering of neutrinos is shown in Fig.~\ref{f:D2nu_unif_vs_GLQ}.  Uniform-density sampling, with each flow representing $2\%$ of the neutrinos, only covers the Fermi-Dirac distribution near its peak, undersampling the low-momentum regime that dominates small-scale clustering.  Interpolation improves the small-scale clustering accuracy and evades the small-scale, high-$u$ numerical noise discussed above, evident in Fig.~\ref{f:D2nu_unif_vs_GLQ}~(top).

\cite{Chen:2020bdf} estimated that, even for the largest $\Mnu$ considered, $\approx 932$~meV, only about a fourth of the neutrinos exhibit significant non-linear clustering.  Allowing for greater accuracy as well as some cosmology-dependence, we choose to emulate the slowest $80\%$ of the flows, reducing our running time and memory usage somewhat.  Increasing this emulated fraction to $100\%$ yields no appreciable improvement in accuracy.  Our final public code, emulating $50$ Gauss-Laguerre quadrature flows, executed on a Dell Optiplex 7010 computer with $24$ threads, runs in $23$ milliseconds and uses $15$~MB of memory.  Reducing the number of flows to $10$ reduces both the running time and the memory usage by $\approx 20\%$.

A further benefit to neglecting the highest-$u$ flows is a reduced sensitivity to numerical errors.  At the largest $k$ and $u$, $\delta(a,k,u)/\dm(a,k) \approx \kfs(a,u)^2 / k^2$ was within a few orders of magnitude of the numerical tolerance of the \fftm{} computations of \cite{Upadhye:2023bgx}. The resulting numerical errors had a negligible impact upon the total $\dnu$.  However, these large-$u$ flows on their own are unreliable.  By extrapolating $\Rnu(a,k,u)$ at large $u$ as in Sec.~\ref{sec:Rnl}, we reduce our sensitivity to these numerical errors.

To summarize, the \enuii{} training set constructed here differs from the \cosmicenu{} training set in two ways.%
\begin{enumerate}
\item \enuii{} individually emulates the $u=0$ Time-RG flow as well as the $40$
  slowest neutrino flows, rather than the momentum deciles of the earlier work.
\item \enuii{} reduces the dynamic range of the training set by dividing by
  the \fastnuf{} linear-response density contrast, as in \EQ{e:Rnu_a_k_u},
  rather than a linear approximation.
\end{enumerate}
Aside from these two, our emulation procedure, using the {\sc sepia} software of \cite{SEPIA}, is identical to that of \cite{Upadhye:2023bgx}, and we refer interested readers to that article for details. Appendix~\ref{app:emuTest} tests our results.

\subsection{Tests of \enuii{}}
\label{sub:emu:enuii}

\begin{figure*}
  \hskip-3mm\includegraphics[width=91mm]{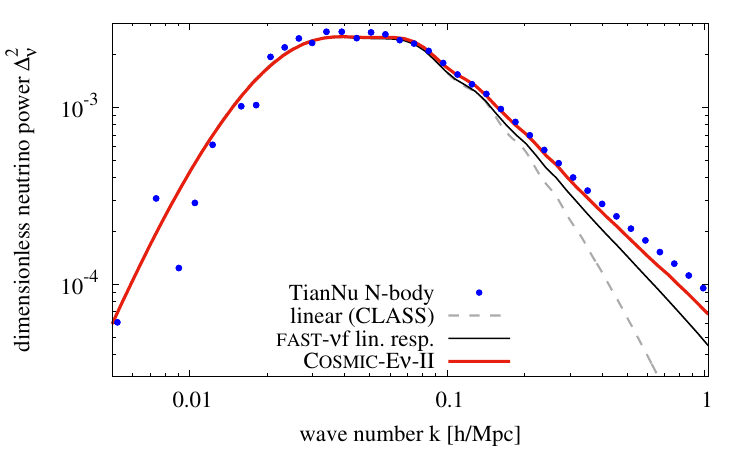}
  \hskip-2mm\includegraphics[width=91mm]{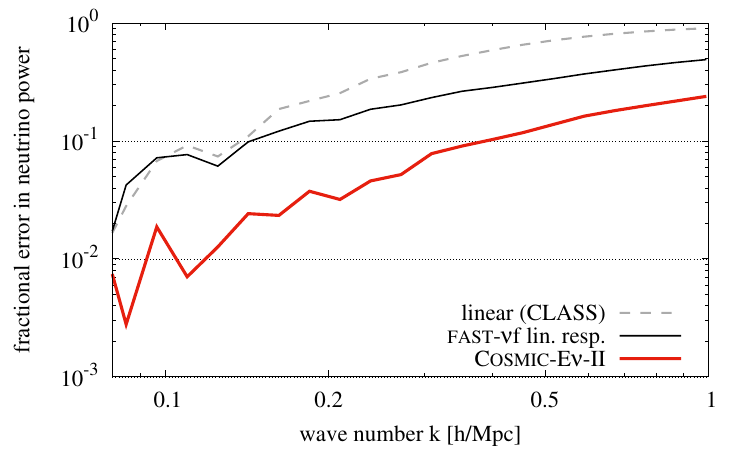}%
  \caption{
    Accuracy of \enuii{} at low neutrino mass.  Shown is the TianNu N-body
    neutrino power spectrum of {\protect \cite{Inman:2016prk}}, with
    $\Mnu = 50$~meV~(NO), compared with: the linear power spectrum of
    \classcode{}; the linear-response power spectrum computed using \fastnuf{};
    and the emulated power spectrum of \enuii{}.
    (Left)~Dimensionless power spectra.
    (Right)~Fractional error in each perturbative power spectrum relative to
    TianNu.
    \label{f:comp_tiannu}
  }
\end{figure*}

We conclude this Section by testing \enuii{} against published N-body power spectra.  The TianNu simulation of \cite{Yu:2016yfe,Emberson:2016ecv,Inman:2016prk} is the highest-resolution neutrino N-body simulation ever performed.  It approximates the minimal-mass normal ordering by a single $50$~meV massive neutrino, with the other two treated as massless.  Figure~\ref{f:comp_tiannu} compares our \fastnuf{} and \enuii{} $\Delta_\nu^2$ against the TianNu power spectrum.

Although noise in the simulation is too large to verify sub-percent-level accuracy for $k \lesssim 0.1~h/$Mpc, Fig.~\ref{f:comp_tiannu}~(Right) confirms that \enuii{} is $1\%-2\%$ accurate around $k=0.1~h/$Mpc.  Errors reach $10\%$ at $k=0.14~h/$Mpc for linear and \fastnuf{} methods, and at $0.40~h/$Mpc for \enuii{}.  At $k=1~h/$Mpc, the error reduces from $90\%$ for linear theory (that is, an order of magnitude) to $49\%$ for \fastnuf{} linear response, to $24\%$ for \enuii{}.  Thus, even for small neutrino masses, non-linearity cannot be neglected, while \enuii{} provides a resonable approximation.

\begin{figure}
  \hskip-3mm\includegraphics[width=92mm]{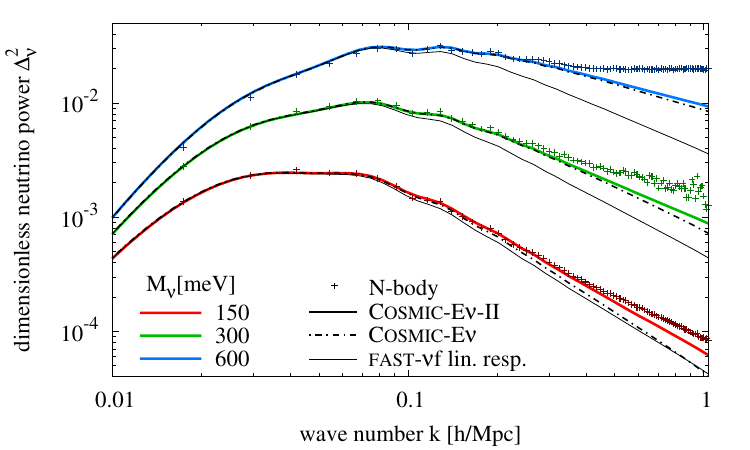}%
  \caption{
    Accuracy of \enuii{} for a range of $\Mnu$.
    Neutrino power spectra of {\protect \cite{Euclid:2022qde}} are compared with
    \enuii{} (thick colored lines), \cosmicenu{} (dot-dashed lines) and
    \fastnuf{} linear response (thin black lines), for the E150DO, E300DO, and
    E600DO models of Table~\ref{t:nLCDM_models}. The E150DO simulation uses the
    SWIFT code, and the other two use Gadget-3.
    \label{f:comp_euclid}
  }
\end{figure}

The Euclid N-body code-comparison project of \cite{Euclid:2022qde} allows us to test the accuracy of \fastnuf{}, \cosmicenu{}, and \enuii{} over a range of $\Mnu$.  At the smallest $\Mnu$, that reference reduces the shot noise using the $\delta f$ method of \cite{Elbers:2020lbn,SWIFT:2023dix}, while at larger masses it uses the Gadget-3 code of \cite{Springel:2005mi,Springel:2008cc}.  Figure~\ref{f:comp_euclid} shows our results.  Evidently, the \enuii{} error $k=1~h/$Mpc exceeds $50\%$ at $\Mnu=600$~meV, though it is significantly more accurate than \fastnuf{} at all scales $k \geq 0.1~h/$Mpc.  At lower $k$, the \fastnuf{}, \cosmicenu{}, and \enuii{} errors reach $10\%$ at $0.13~h/$Mpc, $0.29~h/$Mpc, and $0.36~h/$Mpc, respectively.  Meanwhile, the lower masses highlight the greater accuracy of \enuii{} over \cosmicenu{} due to its improved momentum resolution. Though the two agree up to $k=0.2~h/$Mpc, and are much more accurate than linear response, \cosmicenu{} underestimates high-$k$ power.  For $\Mnu=150$~meV, at $k=1~h/$Mpc, its error exceeds $50\%$, twice that of \enuii{}.

\begin{figure}
  \hskip-5mm\includegraphics[width=92mm]{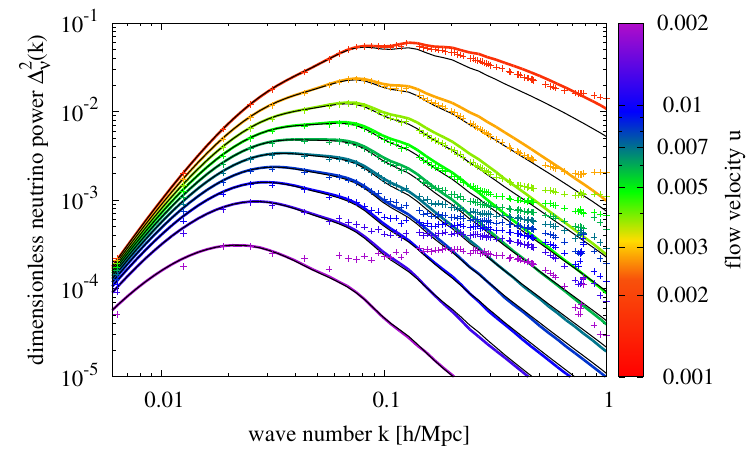}%
  \caption{
    Accuracy of \enuii{} for individual flows.  Neutrino flow power spectra
    of {\protect \cite{Banerjee:2018bxy}} (points) are compared with \enuii{}
    (thick colored lines) and \fastnuf{} linear response (thin black lines) for
    the Q150DO model of Table~\ref{t:nLCDM_models} with $A_s$ raised to
    $2.13\times 10^{-9}$.
    The ``spikes'' in the N-body spectra at $k=0.8~h/$Mpc are grid artifacts
    arising in this method.  For $k$ well below $0.8~h/$Mpc, \enuii{}
    and the simulation agree.
    \label{f:comp_banerjee_150DO}
  }
\end{figure}

N-body simulatons by \cite{Banerjee:2018bxy,Inman:2020oda} also discretize the neutrinos' Fermi-Dirac distribution into momentum flows.  In particular, \cite{Banerjee:2018bxy} presents a simulation increasing the low-momentum resolution, as well as discretizing neutrino directions within each flow in order to reduce the shot noise.  Figure~\ref{f:comp_banerjee_150DO} compares their individual flow power spectra to the predictions of \enuii{}.  Although this N-body method leads to a power spike at $k \approx 0.8~h/$Mpc due to grid artifacts, further studied in \cite{Sullivan:2023ntz}, the two methods agree below this spike.

\begin{figure}
  \hskip-3mm\includegraphics[width=92mm]{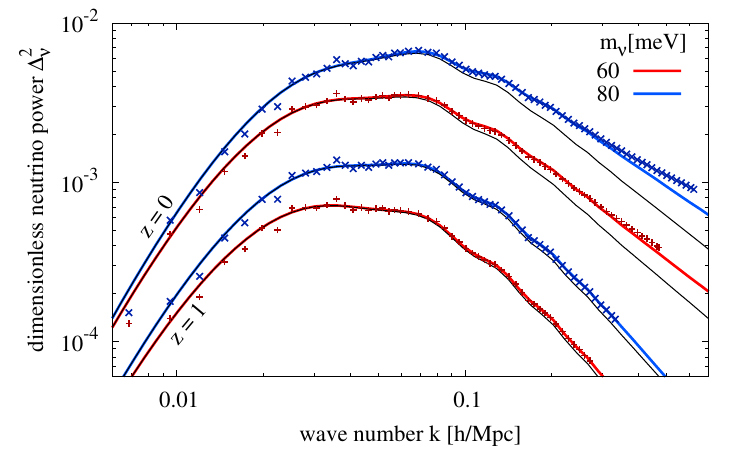}%
  \caption{
    Power spectra for an $80$~meV neutrino and a doubly-degenerate $60$~meV
    neutrino, simulated by {\protect \cite{Adamek:2017uiq}} (points), compared
    with \enuii{} (thick colored lines) and \fastnuf{} linear response (thin
    black lines).  The upper two sets of curves show $z=0$, and the lower two
    $z=1$.  Evidently, \enuii{} accurately recovers the power spectra of
    individual neutrino species, in addition to the total neutrino power tested
    in Figs.~\ref{f:comp_tiannu},~\ref{f:comp_euclid}.
    \label{f:comp_adamek_200NO}
  }
\end{figure}

We conclude this subsection by comparing to the N-body simulation of \cite{Adamek:2017uiq}, which approximated $\Mnu=200$~meV NO neutrinos by simulating two species of mass $60$~meV and one of $80$~meV.  Figure~\ref{f:comp_adamek_200NO} compares \fastnuf{} and \enuii{} to these two simulated power spectra.  In the $k \geq 0.025~h/$Mpc region where the simulation noise has died down, \enuii{} at $z=0$ is accurate to $<10\%$ up to $k=0.5~h/$Mpc, the smallest scale shown, for $60$~meV neutrinos, and to $k=0.4~h/$Mpc for $80$~meV neutrinos.  For \fastnuf{} linear response, these upper limits are $0.13~h/$Mpc and $0.12~h/$Mpc, respectively.  At $z=1$, \enuii{} is accurate to better than $1\%$ at the highest $k$ shown, $\approx 0.3~h/$Mpc, while linear response underestimates power by $8\%$ for $60$~meV and $12\%$ for $80$~meV.  Thus, we see that \enuii{} accurately computes the neutrino power spectrum up to $k \approx 0.4~h/$Mpc over a broad range of masses, and approximates it to reasonable accuracy up to $k=1~h/$Mpc for low masses.

\section{Painting neutrinos onto halos}
\label{sec:halo}


Neutrino halos around cb halos typically extend out to about eight times the virial radius $\Rvir$ of the cb halo, as shown by \cite{LoVerde:2013lta}.  Since \enuii{} is ultimately based upon non-linear perturbation theory, we cannot realistically hope to predict the inner neutrino halo profile, $r \lesssim \Rvir$.  However, we should be able to predict the neutrino density in the halo outskirts, $\Rvir \lesssim r \lesssim 10 \Rvir$, where much of the neutrino population is to be found.  We further simplify our task by focusing on very massive cb halos, $M = 10^{15}M_\odot/h$, which are large enough to dominate their local enviromnents, and rare enough to be separated from one another and larger halos by $\sim 100$~Mpc$/h$.

For such halos, the Quijote simulation suite of \cite{Villaescusa-Navarro:2019bje} is ideal.  Its cosmological parameters are listed as series Q in Table~\ref{t:nLCDM_models}, and $500$  simulations are available for each of the neutrino mass sums $100$~meV, $200$~meV, and $400$~meV.  For these models, $M=10^{15}M_\odot/h$ corresponds to $\Rvir \approx 2.5~{\rm Mpc}/h$.  Each such Quijote halo has $\approx 1500$ particles, allowing its density profile to be resolved down to distances a few times its force-softening scale of $R_{\rm soft} = 50~{\rm kpc}/h$.

Our approach to modelling the neutrino halo is to ``paint'' neutrinos onto the cb overdensity in Fourier space:
\begin{equation}
  \delta_{\nu,\mathrm{halo}}(k)
  \approx \delta_\mathrm{cb, halo}(k) \sqrt{\frac{\DDnu(k)}{\DDcb(k)}}.
  \label{e:nu_paint}
\end{equation}
Thus we begin by finding the cb density profile in the halo and its outskirts.  We make no distinction between bound and unbound particles, but include all particles within a given distance of each Quijote FoF halo. The transition between the one-halo and two-halo cb density components occurs in the outskirts region, making it theoretically challenging even for the cb halo.  Although the density profile ought to be well-described by linear perturbation theory far from the halo, the profile in the outskirts is difficult to predict from first principles, leading \cite{Diemer:2014xya,Diemer:2017bwl} to suggest a decaying power law fit with a central core.

We thus model the cb density in the outskirts as a function proportional to $r^{-\nOut}$ convolved with a Lorenzian function to ensure an inner core, parameterizing it in Fourier space as $\rho_{\rm cb,O}(k)/M = \COut (k\Rvir)^{\nOut-3} \exp(-\lvir k \Rvir)$.  The inner density is well-described by the profile of \cite{Navarro:1996gj} (NFW) with variable concentration $c_{\rm I}$, which we suppress at large scales with the factor $\exp(-r^2/(\gvir \Rvir^2))$, resulting in five free parameters ($c_{\rm I}$, $\gvir$, $\lvir$, $\COut$, $\nOut$) for halos of a given mass $M$.

\begin{figure}
  \hskip-3mm\includegraphics[width=90mm]{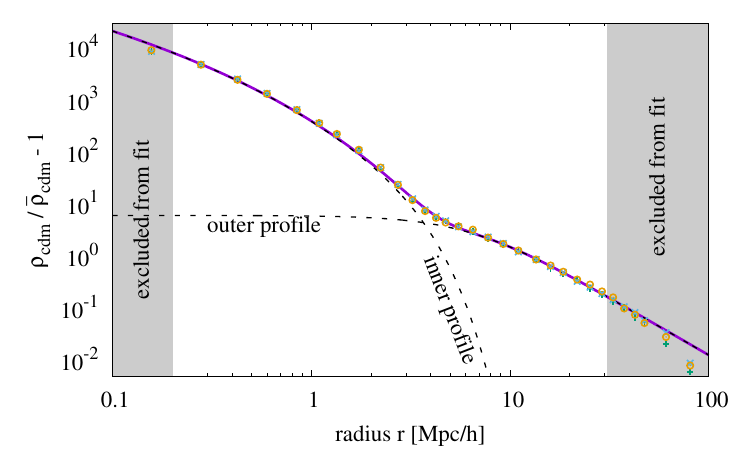}%
  \caption{
    NFW+power law fit to the density profiles of $M=10^{15}M_\odot/h$ cb halos.
    The solid line shows our fit, with parameters given in the text, while
    the three sets of point represent three different Quijote simulations.
    The two agree well in the halo outskirts,
    $\Rvir = 2.5~{\rm Mpc}/h \lesssim r \lesssim 10 \Rvir = 25~{\rm Mpc}/h$.
    \label{f:cb_halo_fit_quijote}
  }
\end{figure}

Choosing all halos within $10\%$ of $10^{15}M_\odot/h$ in three different $\Mnu=400$~meV Quijote runs, and fitting over the range $4 R_{\rm soft} = 0.2~{\rm Mpc}/h \leq r \leq 31~{\rm Mpc}/h$, we find $c_{\rm I}=3.9$, $\gvir=1.6$, $\lvir=2.6$, $\COut=1.2$, and $\nOut=2.2$.  We find similar numbers for the other $\Mnu$, using $12$ and $25$ simulations, respectively, for $\Mnu=200$~meV and $100$~meV.  This concentration $c_{\rm I} \approx 4$ is reasonable for the most massive halos, while $\nOut < 3$ falls off more slowly than the NFW profile, consistent with large halos forming in overdense environments.  Figure~\ref{f:cb_halo_fit_quijote} shows our result, which agrees in the halo outskirts.  Note that the N-body density begins to fall below our fit beyond about three times the Lagrangian radius $R_{\rm L} = (18\pi^2)^{1/3}\Rvir \approx 14~{\rm Mpc}/h$, suggesting that these halos live in overdense regions extending to a few $R_{\rm L}$.

\begin{figure}
  \hskip-2mm\includegraphics[width=87mm]{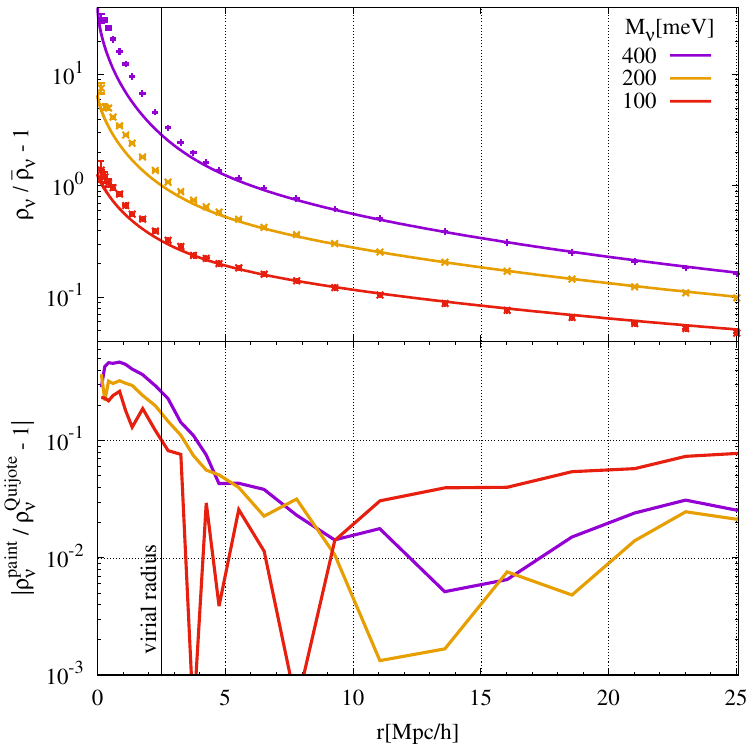}%
  \caption{
    Neutrino density profiles painted onto Quijote simulations using
    \EQ{e:nu_paint} (lines), compared with the actual Quijote neutrino
    profiles (points).
    (Top)~Neutrino density contrast.
    (Bottom)~Fractional error between painted and Quijote profiles.  Our simple,
    parameter-free painting procedure accurately matches the N-body density
    profile outside of $1-2$~virial radii.
    \label{f:nu_halo_quijote}
  }
\end{figure}

Finally, we use \enuii{} to paint neutrinos onto the cb halo profile.  Since the emulator is limited to $k \leq 1~h/$Mpc, we use logarithmic extrapolation to approximate the power spectrum at larger $k$. We verify that this has a negligible effect upon the density profile except in the inner megaparsec.

Figure~\ref{f:nu_halo_quijote} shows our results for three different neutrino masses.  We have included all neutrinos around all cb halos within $10\%$ of $10^{15}M_\odot/h$, and have used $3$, $12$, and $25$ simulations, respectively, for $\Mnu$ of $400$~meV, $200$~meV, and $100$~meV.   For all three masses, errors in the $2\Rvir \leq r \leq 10\Rvir$ region are $< 8\%$, demonstrating that our fast, parameter-free \enuii{}-based painting procedure of \EQ{e:nu_paint} is a reasonable approximation.

Not surprisingly, errors rise to $30\%$-$50\%$ in the inner $1-2$ megaparsecs, reflecting the breakdown of perturbation theory.  Accurate prediction of the neutrino density profile in the inner $2\Rvir$ would likely require a neutrino halo model, a fascinating open problem which is beyond the scope of the present article.  Nevertheless, our ability to predict the neutrino profile in the halo outskirts from the simple, parameter-free painting of \EQ{e:nu_paint} is encouraging.

\begin{figure}
  \hskip-3mm\includegraphics[width=90mm]{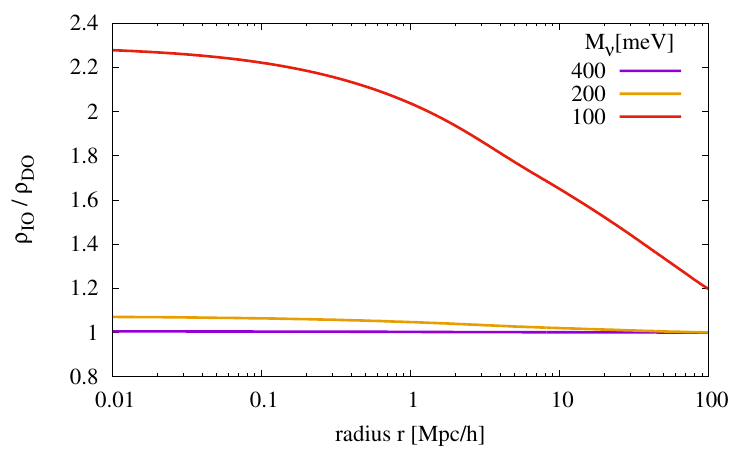}%
  \caption{
    Ratio of neutrino density profiles assuming IO vs.~DO mass orderings.
    Neutrinos are painted onto the $M=10^{15}~M_\odot/h$ halo profile shown in
    Fig.~\ref{f:cb_halo_fit_quijote}.  Evidently, near the lower-$\Mnu$ limit
    of IO neutrinos, they cluster far more strongly than DO neutrinos of the
    same $\Mnu$.
    \label{f:rho_quijote_IO_vs_DO}
  }
\end{figure}

Finally, we consider the hierarchy-dependence of the neutrino profile.  Suppose, following Fig.~\ref{f:nu_halo_quijote}, that we regard halo painting for $\Mnu=100$~meV as $20\%-30\%$ accurate within the virial radius and $10\%$ outside of it.  Then we may ask, how differently would neutrinos with IO masses cluster around this halo?  Figure~\ref{f:rho_quijote_IO_vs_DO} shows the result, along with the other two Quijote neutrino masses.  Linear theory predicts an IO-to-DO ratio of $(50~{\rm meV})^2/(33.{\bar 3}~{\rm meV})^2 = 9/4=2.25$ at $\Mnu=100$~meV, if we approximate IO in this case as two $50$~meV neutrinos, while non-linear theory predicts a small enhancement; see \cite{Ringwald:2004np,Upadhye:2023bgx}.  This is consistent with the figure, which shows a density ratio of $2.29$ at low $r$, $1.9$ at $\Rvir$, and $1.8$ at $2\Rvir$.

\section{Conclusions}
\label{sec:con}

This article has introduced and combined two innovations relevant to the cosmological clustering of massive neutrinos.  Firstly, in Sec.~\ref{sec:fnf}, we have developed the \fastnuf{} method for rapidly computing the linear density contrasts of individual neutrino thermal-velocity flows, given any linear or non-linear matter density contrast.  It is applicable to fully linear calculations after the neutrinos have become non-relativistic, as well as to linear response calculations if coupled to a non-linear matter clustering calculation.

In addition to the total neutrino power, it rapidly provides high-resolution predictions of the thermal-velocity-dependence of neutrino clustering.  Moreover, since the multi-flow description is applicable to any hot dark matter, given an appropriately-chosen distribution function, \fastnuf{} can be used to study non-standard models such as axions and non-thermal neutrinos.

Secondly, in Sec.~\ref{sec:emu}, we used \fastnuf{} to construct a new thermal-momentum-space emulator for the neutrino density contrast, using the existing \cosmicenu{} emulator training data. By emulating the non-linear enhancement ratio $\Rnu(a,k,u)$ of \EQ{e:Rnu_a_k_u}, that is, the ratio of non-linear and linear-response perturbation theories, we reduced by orders of magnitude the dynamic range of the previous training set.  This reduced dynamic range makes $\Rnu$ amenable to interpolation, which allowed us to remedy the main limitation of \cosmicenu{}, its limited momentum resolution, thereby improving the small-scale, small-mass clustering predictions of \enuii{} relative to the previous emulator by a factor greater than two.

Further, we demonstrated that the previous degenerate-neutrino-mass emulator could be generalized to the normal and inverted mass orderings.  As the cosmological data come to prefer lower $\Mnu$, the distinction between these grows increasingly interesting.  The success of our approach also suggests its generalization to non-standard hot dark matter species, provided that their masses and temperatures are not significantly different from those of neutrinos, although this would require further testing and calibration.

After verifying that \enuii{} power spectra matched the underlying non-linear perturbation theory as well as N-body neutrino simulations, we applied it to predicting the clustering of neutrinos outside large dark matter halos.  Simply by painting neutrinos onto the halos in Fourier space, we predicted the neutrino density profiles between two and ten virial radii to better than $10\%$.  As knowledge of the boundaries and outskirts of halos improves, calculations such as ours could prove invaluable for developing new observables providing cross-checks on neutrino mass measurements from larger scales.

\subsection*{Acknowledgments}

We are grateful to J.~Kwan, X.~Liu, I.~G.~McCarthy, M.~Mosbech, and Y.~Y.~Y.~Wong for insightful conversations during the progress of this work.  AU acknowledges the support of the NSFC under grant No.\ 12473005, and the Yunnan Key Laboratory of Survey Science, Project No.\ 202449CE340002. YL is supported by the Major Key Project of Peng Cheng Laboratory and the National Key Research and Development Program of China under grant No.\ 2023YFA1605600.

\subsection*{Data availability}

No new experimental data were generated or analyzed in support of this research. The \fastnuf{} and \enuii{} codes developed in this article are respectively available online at {\tt http://codeberg.org/upadhye/FASTnuf} and {\tt http://codeberg.org/upadhye/Cosmic-Enu-II}~.

\bibliographystyle{mnras}
\bibliography{enuii}

\appendix

\section{Cubic interpolation}
\label{app:gcubic}

We list here the quantities necessary for applying the \fastnuf{} procedure with $N_g=4$ powers of $\Hco s$ in the definition of $g$, \EQ{e:def_g_s_k}, and its interpolation using cubic polynomials.  Firstly, defining $\Iys_n(y) = \int_y^\infty dy' \cos(y') / (y')^n$ and $\Iyc = \int_0^y dy' \sin(y') / (y')^n$ as in \EQ{e:def_Iys_Iyc},%
\begin{eqnarray}
  \Iys_{1,j} &=& -\Ci(y_j), \quad \Iyc_{1,j} = \Si(y_j),
  \label{e:Iy1}
  \\
  \Iys_{2,j} &=& \Si(y_{j}) + \frac{\cos(y_{j})}{y_{j}},
  \quad
  \Iyc_{2,j} = \Ci(y_{j}) - \frac{\sin(y_{j})}{y_{j}},
  \label{e:Iy2}
  \\
  \Iys_{3,j} &=&
  \frac{\Ci(y_{j})}{2} - \frac{\sin(y_{j})}{2y_{j}}
  + \frac{\cos(y_{j})}{2(y_{j})^2},
  \label{e:Iys3}
  \\
  \Iyc_{3,j} &=&
  -\frac{\Si(y_{j})}{2} - \frac{\cos(y_{j})}{2y_{j}}
  - \frac{\sin(y_{j})}{2(y_{j})^2},
  \label{e:Iyc3}
  \\
  \Iys_{4,j} &=&
  -\frac{\Si(y_{j})}{6} - \frac{\cos(y_{j})}{6y_{j}}
  - \frac{\sin(y_{j})}{6(y_{j})^2}
  + \frac{\cos(y_{j})}{3(y_{j})^3},
  \label{e:Iys4}
  \\
  \Iyc_{4,j} &=&
  -\frac{\Ci(y_{j})}{6} + \frac{\sin(y_{j})}{6y_{j}}
  - \frac{\cos(y_{j})}{6(y_{j})^2}
  - \frac{\sin(y_{j})}{3(y_{j})^3}~.
  \label{e:Iyc4}
\end{eqnarray}

Secondly, we show how to compute the coefficients $g_n$, $0 \leq n \leq 3$, interpolating some function $g(x)$ at the non-zero points $x_0$, $x_1$, $x_2$, and $x_3$.  The $m$-th interpolating polynomial $L_m(x):= \prod_{p \neq m} (x-x_p)/(x_m-x_p)$ is defined to be one at $x_m$ and zero at all $x_p$ for $p\neq m$, so the polynomial interpolating $g(x)$ at these points is $\sum_m L_m(x) g(x_m)$.  Each $L_m(x)$ can itself be expressed as $L_m(x) = \sum_n L_{mn} x^n$, where, for cubic polynomials,
\begin{eqnarray}
  L_{m,3} &=& \prod_{p\neq m} \frac{1}{x_m-x_p},
  \quad
  \frac{L_{m,2}}{L_{m,3}} = x_m - S^{(+)}
  \\
  \frac{L_{m,0}}{L_{m,3}} &=& -\frac{P}{x_m},
  \qquad\qquad
  \frac{L_{m,1}}{L_{m,0}} = x_m^{-1} - S^{(-)},
  \\
  P &:=& \prod_{m=0}^3 x_m,
  \qquad\quad~
  S^{(\pm)} := \sum_{m=0}^3 x^{\pm 1}_m.
\end{eqnarray}
Thus, finally, we have
\begin{equation}
  g(x)\approx \sum_{n=0}^3 g_n x^n
  ~~\mathrm{where}~~
  g_n = \sum_{m=0}^3 g(x_m) L_{mn}.
  \label{e:g_n_cubic}
\end{equation}
For given wave number $k$, and superconformal time interval $[\Hco s_j, \Hco s_{j+1}] \subset [x_0,x_3]$, the $g_n$ of \EQ{e:g_n_cubic} are precisely the $g_{nj}(k)$ of Sec.~\ref{sub:fnf:gen_cos}.  Where possible, we choose $x_m = \Hco s_{j+m-1}$.

\section{Holdout and out-of-sample tests}
\label{app:emuTest}

\begin{figure}
  \hskip-3mm\includegraphics[width=89mm]{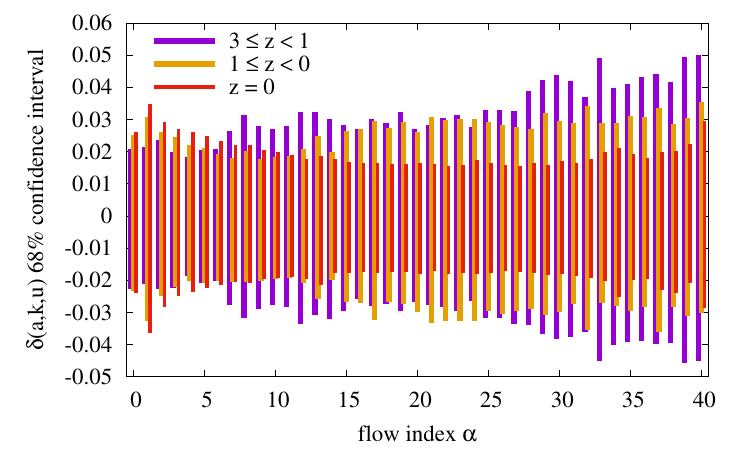}%
  \caption{
    Holdout tests for individual flows.  The vertical axis shows the $68\%$
    confidence region for each flow, maximized over all $k$ between
    $10^{-3}~h/$Mpc and $1~h/$Mpc, and all $z$ in the ranges shown.
    \label{f:cross-val_flows}
  }
\end{figure}

\begin{figure}
  \hskip-2mm\includegraphics[width=87mm]{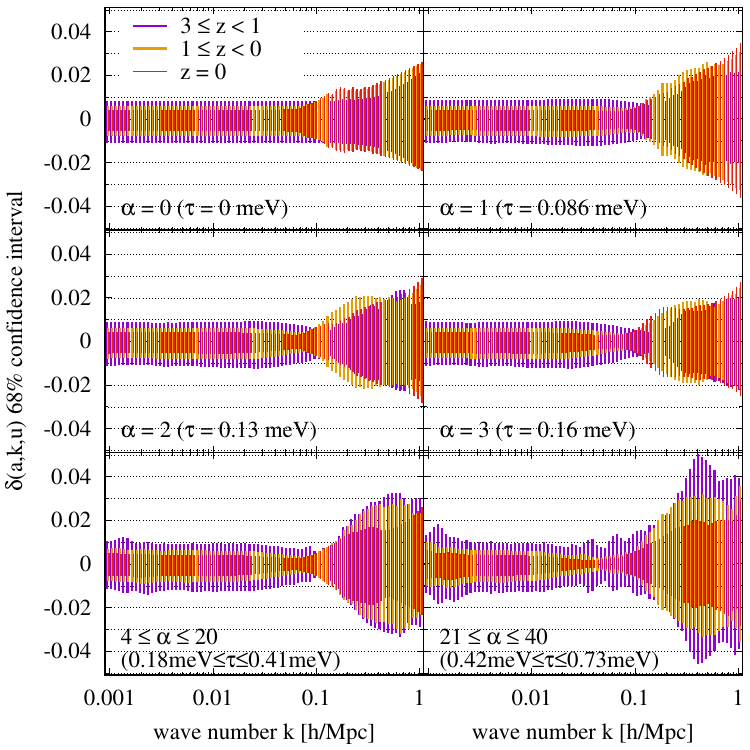}%
  \caption{
    Holdout tests for individual wave numbers.  The vertical axis shows the
    $68\%$ confidence region for each flow, maximized over all redshifts $z$ and
    all flow velocity indices $\alpha$ in the ranges shown.  Each of the first
    four flows is labelled by its respective flow momentum $\tau = m_\nu u$,
    while the remaining flows are combined into $\tau$ ranges.
    \label{f:cross-val_k}
  }
\end{figure}

Cross-validation is a technique for placing an upper bound on the emulator error using its existing training set.  For each of the training models, we compare the training data to the predictions of a new emulator which is built using every other model; thus, these are referred to as ``holdout tests.''  Since holding out a single model creates a gap in the training set at that point, holdout tests should overestimate the error by an amount which \cite{Upadhye:2023bgx} estimated to be $60\%$ for the earlier \cosmicenu{} emulator.

Figure~\ref{f:cross-val_flows} shows errors from holdout testing for each individual flow emulated, with $\alpha=0$ ($u=0$) corresponding to Time-RG, and each subsequent flow representing $2\%$ of the neutrino number density.  At low $z$, the larger non-linear corrections of the slower flows, hence the larger dynamic range of $\Rnu$, makes emulation more difficult, which is reflected in the larger errors.  The fastest flows, especially at high redshifts, cluster much more weakly than cold matter, leading to greater numerical errors and increased emulation difficulty.  Across all flows, the error is consistenly $\leq 5\%$ between redshifts of one and three, and $< 4\%$ below redshift one.

Figure~\ref{f:cross-val_k} separates the cross-validation errors by the wave number.  Evidently, errors are $<2\%$ on large scales, $k \leq 0.1~h/$Mpc, and increase on smaller scales.  Moreover, the relative accuracy of the slow flows, $\approx 3.5\%$ for $\alpha=1$ and and $<3\%$ for $\alpha$ of $0$, $2$, and $3$, imply accurate small-scale power spectrum predictions, as these flows dominate small-scale clustering.

\begin{table}
  \caption{
    Out-of-sample (OOS) models used for emulation tests; referred to as
    OOS\# followed by the number in the first column.
    \label{t:oos_models}
  }
  \renewcommand{\tabcolsep}{0.75mm}
  \begin{footnotesize}
    \begin{tabular}{ccccccccc}
      \# & $\Omo h^2$ & $\Obo h^2$ & $\Ono h^2$ & $\sigma_8$
      & $h$ & $n_{\rm s}$ & $w_0$ & $w_a$\\
      \hline
      $01$ & $0.1433$ & $0.02228$ & $0.008078$ & $0.8389$
      & $0.7822$ & $0.9667$ & $-0.8000$ & $-0.0111$
      \\
      $02$ & $0.1333 $ & $0.02170 $ & $0.005311 $ & $0.8233 $
      & $0.7444 $ & $0.9778 $ & $-1.1560 $ & $-1.1220 $
      \\
      $03$ & $0.1450 $ & $0.02184 $ & $0.003467 $ & $0.8078 $
      & $0.6689 $ & $0.9000 $ & $-0.9333 $ & $- 0.5667 $
      \\
      $04$ & $0.1367 $ & $0.02271 $ & $0.002544 $ & $0.8544 $
      & $0.8200 $ & $0.9444 $ & $-0.8889 $ & $- 1.4000 $
      \\
      $05$ & $0.1400 $ & $0.02257 $ & $0.009000 $ & $0.7300 $
      & $ 0.7067$ & $0.9889 $ & $-0.9778 $ & $- 0.8444 $
      \\
      $06$ & $0.1350 $ & $0.02213 $ & $0.000700 $ & $0.8700 $
      & $0.7633 $ & $0.9111 $ & $-1.0220 $ & $0.5444 $
      \\
      $07$ & $0.1383 $ & $0.02199 $ & $0.007156 $ & $0.7456 $
      & $0.6500 $ & $0.9556 $ & $-1.1110 $ & $1.1000 $
      \\
      $08$ & $ 0.1300$ & $0.02286 $ & $0.006233 $ & $0.7922 $
      & $0.8011 $ & $1.0000 $ & $-1.0670 $ & $0.2667 $
      \\
      $09$ & $0.1417 $ & $0.02300 $ & $0.004389 $ & $0.7767 $
      & $0.7256 $ & $0.9222 $ & $-0.8444 $ & $0.8222 $
      \\
      $10$ & $0.1317 $ & $0.02242 $ & $0.001622 $ & $0.7611 $
      & $0.6878 $ & $0.9333 $ & $-1.2000 $ & $- 0.2889 $
      \\
    \end{tabular}
  \end{footnotesize}
\end{table}

\begin{figure}
  \hskip-3mm\includegraphics[width=90mm]{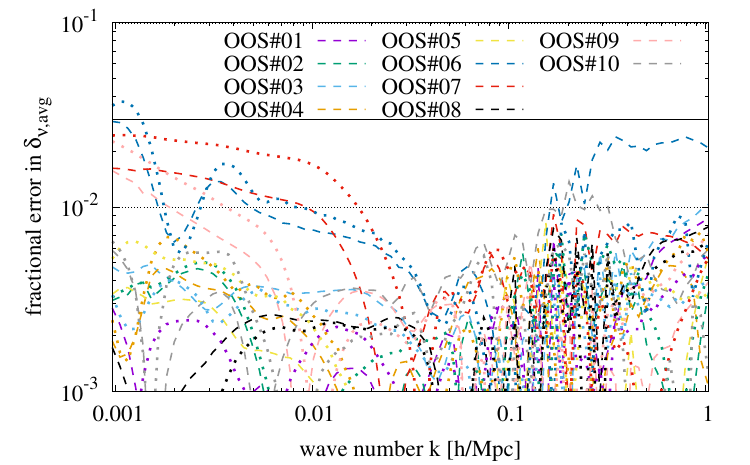}%
  \caption{
    Out-of-sample tests comparing \enuii{} to \fftm-II for the models of
    Table~\ref{t:oos_models}, at $z=0$ (dashed) and $z=1$ (dotted).
    Thin dotted and solid horizontal lines show $1\%$ and $3\%$, respectively.
    We compare the neutrino density contrast
    averaged over the $40$ emulated neutrino flows, representing the
    slowest $80\%$ of the neutrinos.
    \label{f:oos_dnuAvg}
  }
\end{figure}

Next, we proceed to out-of-sample tests comparing \enuii{} to independent runs of \fftm{}, for the models shown in Table~\ref{t:oos_models}.  Figure~\ref{f:oos_dnuAvg} begins by testing the neutrino density contrast averaged over all emulated neutrino flows.  At both redshifts and at all wave numbers, $70\%$ of the out-of-sample models are accurate to $<1\%$.  All ten models are correct to $<3\%$ ($<4\%$) at $z=0$ ($z=1$).  At small scales, the least accurate model, with an error $\gtrsim 2\%$, is OOS\#06, which has a neutrino mass sum $\Mnu = 65.3$~meV near the lower bound of \cite{Esteban:2024eli}.

\begin{figure}
  \hskip-3mm\includegraphics[width=87mm]{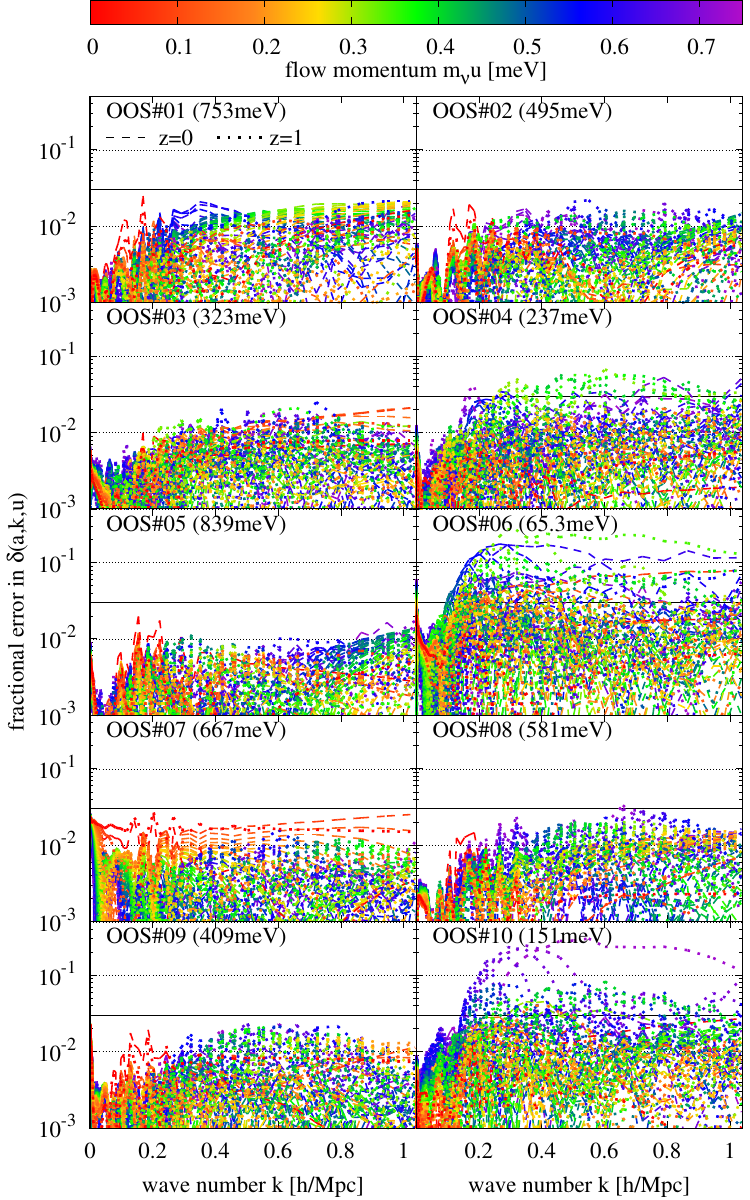}%
  \caption{
    Out-of-sample tests comparing \enuii{} to \fftm-II for the models of
    Table~\ref{t:oos_models}, at $z=0$ (dashed) and $z=1$ (dotted).
    We compare the $u=0$ Time-RG flows as well as the $40$ slowest
    $\nu$ flows, each corresponding to $2\%$ of the neutrino density.
    Solid horizontal lines represents $3\%$ error.
    \label{f:oos_flows}
  }
\end{figure}

Figure~\ref{f:oos_flows} provides greater insight into the momentum-dependence of these errors.  Considering both redshifts, all flows, and all wave numbers, we see that for $70\%$ of the models, \enuii{} accurately predicts the density contrast to $\leq 3.2\%$.  The three least-accurate models, OOS\#06, OOS\#10, and OOS\#04, are precisely those with the smallest neutrino masses.  At $z=0$, two of these three have errors $\leq 5\%$ for all flows and wave numbers, while the remaining model, OOS\#06, has errors of up to $17\%$ for the fast flows.

\label{lastpage}
\end{document}